\documentclass[10pt,twocolumn,superscriptaddress,english,10pt,prl,%linenumbers,showpacs,
floatfix,aps%,preprint
]{revtex4-2}

\usepackage[english]{babel}
\usepackage[utf8]{inputenc}
\usepackage[T1]{fontenc}

\usepackage{amsmath}
\usepackage{physics}

\usepackage{graphicx}
\usepackage{microtype}
\usepackage{xcolor}

\newcommand\dIdV{$\dv*{I}{V}$}
\newcommand\mos{MoS\textsubscript{2}}
\newcommand\cfth{CF\textsubscript{3}-3P-SH}
\newcommand\cft{CF\textsubscript{3}-3P-S}

\usepackage[backref=none,bookmarksnumbered=true,bookmarks=true,  	bookmarksopen=true,colorlinks=true,citecolor=blue,linkcolor=blue, 	anchorcolor=green,urlcolor=blue,unicode=false]{hyperref}

\begin{document}

\title{An atomic-scale perspective on individual thiol-terminated molecules anchored to single S vacancies in \mos}%

\author{J.\,Rika Simon}
\affiliation{\mbox{Fachbereich Physik, Freie Universit\"at Berlin, 14195 Berlin, Germany}}

\author{Dmitrii Maksimov}
\affiliation{Max Planck Institute for the Structure and Dynamics of Matter, 22761 Hamburg, Germany}

\author{Christian Lotze}
\affiliation{\mbox{Fachbereich Physik, Freie Universit\"at Berlin, 14195 Berlin, Germany}}

\author{Paul Wiechers}
\affiliation{\mbox{Fachbereich Physik, Freie Universit\"at Berlin, 14195 Berlin, Germany}}

\author{Juan~Pablo~Guerrero~Felipe}
\affiliation{\mbox{Fachbereich Physik, Freie Universit\"at Berlin, 14195 Berlin, Germany}}

\author{Björn Kobin}
\affiliation{Institut für Chemie, Humboldt-Universität zu Berlin, 12489 Berlin, Germany}

\author{Jutta Schwarz}
\affiliation{Institut für Chemie, Humboldt-Universität zu Berlin, 12489 Berlin, Germany}

\author{Stefan Hecht}
\affiliation{Institut für Chemie, Humboldt-Universität zu Berlin, 12489 Berlin, Germany}
\affiliation{Center for Science of Materials, 12489 Berlin, Germany}

\author{Katharina J. Franke}
\affiliation{\mbox{Fachbereich Physik, Freie Universit\"at Berlin, 14195 Berlin, Germany}}
\email{franke@physik.fu-berlin.de}

\author{Mariana Rossi}
\affiliation{Max Planck Institute for the Structure and Dynamics of Matter, 22761 Hamburg, Germany}
\email{mariana.rossi@mpsd.mpg.de}

\date{\today}

\begin{abstract}

Sulphur vacancies in \mos\ on Au(111) have been shown to be negatively charged as reflected by a Kondo resonance. Here, we use scanning tunneling microscopy to show that these vacancies serve as anchoring sites for thiol-based molecules (\cfth) with two distinct reaction products, one of them showing a Kondo resonance. Based on comparisons with density-functional theory (DFT) calculations, including a random structure search and computation of energies and electronic properties at a hybrid exchange-correlation functional level, we conclude that both anchored molecules are charge neutral. One of them is an anchored intact \cfth\ molecule while the other one is the result of catalytically activated dehydrogenation to \cft\ with subsequent anchoring. Our investigations highlight a perspective of functionalizing defects with thiol-terminated molecules that can be equipped with additional functional groups, such as charge donor- or acceptor-moieties, switching units or magnetic centers.  

\end{abstract}

\maketitle

\section{Introduction}

Transition metal dichalcogenides (TMDCs) are a class of materials with unwavering popularity for the last 50~years~\cite{Wilson1969}. The recent interest in these materials originates from the stunning possibilities of isolating single monolayers of these materials and stacking them in almost all imaginable ways, i.\,e.\ combinations of different materials, stacking order and rotation of the layers~\cite{Novoselov2016}. Within the stacks, a single layer is already distinct from its bulk counterpart in that the electronic bandgap is changed due to confinement effects, e.\,g.\ transforming an indirect bandgap to a direct one in MoS$_2$~\cite{Mak2010,Splendiani2010}.

The tunability of electronic properties through variable stacking allows for versatile applications in optoelectronics~\cite{Amani2015}, in field effect transistors~\cite{Radisavljevic2011}, in nanoelectromechanical systems~\cite{Duerloo2012} and in spintronics~\cite{Wang2012}. The 2D nature does not only offer unprecedented flexibility, but also comes along with opportunities of device miniaturization~\cite{Joksas2022,Salahuddin2018}.

Owing to the 2D nature, a very dilute amount of defects already has a major effect on the electronic properties of the material. Often they introduce localized states inside the bandgap of a semiconducting monolayer which may be detrimental for certain applications but also beneficial for others~\cite{Yu2014, Nan2014, Lin2016, Liang2021, Chee2020}.
Attaining control over defects opens one more channel for tuning their local and global electronic properties -- an area usually coined as defect engineering.
The tunability can be enhanced even further by the functionalization of defects with molecular adsorbates~\cite{Huang2018,Bertolazzi2018, Daukiya2019}. This functionalization can lead to further possible charge transfer processes~\cite{Park2019} and doping~\cite{Gali2021}. The hybrid systems may act as biosensors~\cite{Li2012,FathiHafshejani2021}, and offer specific reaction sites~\cite{Asadi2016}. 

Despite these fascinating opportunities, an atomistic understanding of the molecular functionalization in terms of the precise molecular configuration, the character of their bonding to the TMDC substrate and the resulting local electronic structure is largely in its infancy. In this paper, we study a molecular model system based on a thiol-terminated organic molecule and investigate its bonding properties to S vacancies in \mos. By combining scanning tunneling microscopy and theoretical modeling we gain fundamental insights into the atomic-scale bonding and electronic properties. We surmise that the understanding of this simple model system will form a basis for the design of future hybrid systems with more specific functions.

\begin{figure*}
  \centering
\includegraphics[width=0.9\linewidth]{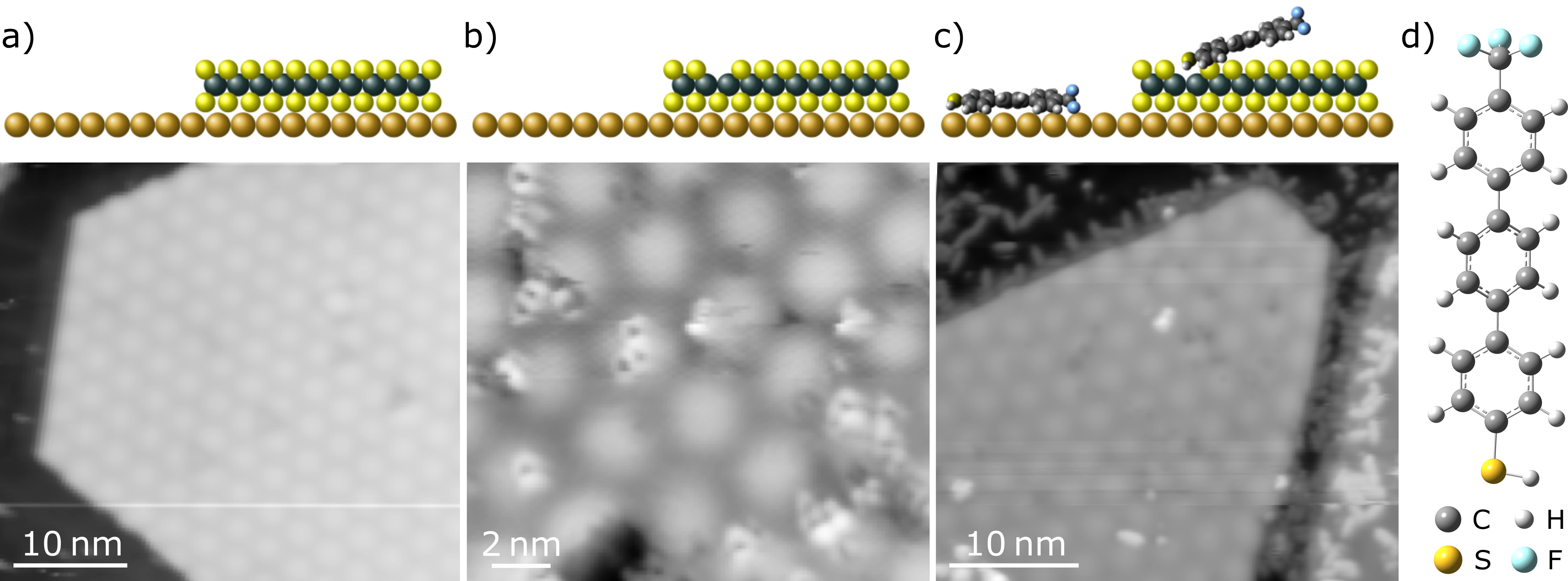}
\caption{Overview STM topographies and sketches of the preparation steps: a) \mos\ on Au(111), showing the characteristic moiré pattern of the \mos\ island (Setpoint: 2\,V, 200\,pA).
  b) Top-layer sulphur vacancies in \mos/Au(111), visible as triangular shapes, created by sputtering with Ne ions (setpoint: 50\,mV, 200\,pA).
  c) \cft(H) on \mos/Au(111) after evaporation and subsequent annealing. Most \cft(H) molecules cluster together on the Au(111) surface, with only a few remaining on the \mos\ island (setpoint: 2\,V, 30\,pA).
d) Molecular model of \cfth\ in gas-phase (grey: carbon, white: hydrogen, blue: fluorine, yellow: sulphur).}
\label{fig:prep_overview}
\end{figure*}

In more detail, we create S vacancies in the terminating S layer of \mos\ on Au(111). The interaction of \mos\ with the Au substrate has been shown to create negatively charged vacancies that localize a single electron at low temperatures~\cite{Trishin2023b}. Nevertheless, an analysis of defect formation energies suggests that a fraction of neutral vacancies can be present at higher temperatures~\cite{Akkoush2023,komsaprb2015}. In both scenarios, these sites are expected to be highly reactive and ideal candidates for anchoring molecules. We expose these defects to 4''-(tri\-fluoro\-methyl)\--[1,1':4',1''-terphenyl]-4-thiol (\cfth) molecules (Fig.\,\ref{fig:prep_overview}d). Thiol-based molecules are prone to covalently bind to S vacancies~\cite{Cho2015,Gali2021,LiAngew2017} and passivate them. In line with these expectations, we observe an anchoring of \cfth\ molecules to S defects on \mos/Au(111) and track the resulting changes in electronic structure in tunneling spectra. 

We rationalize the resulting atomic and electronic structure by density-functional theory simulations. We find robust hybrid molecule-substrate localized electronic states and discuss the possibility of vacancies inducing a dehydrogenation of the molecule to \cft. We suggest that both the charged and neutral state of the S vacancy may play a role in the anchoring mechanism. However, regardless of the initial charge state of the vacancy, and whether dehydrogenation happens or not, a neutral anchored molecule is found to be the most stable state. This observation is compatible with two distinct experimentally observed molecular states, one of them exhibiting a Kondo resonance.

\section{Experimental and theoretical details}
The Au(111) surface was prepared for the \mos\ growth by repeated cleaning cycles of sputtering and annealing under ultra-high vacuum (UHV) conditions. A monolayer of \mos\ was then grown on the atomically clean Au(111) substrate by depositing Mo atoms in an H\textsubscript{2}S-atmosphere of $1 \times 10^{-5}$\,mbar onto the sample kept at 820\,K~\cite{Gronborg2015, Krane2018a}. The as-prepared sample was inspected in the STM to assure the successful formation of monolayer-islands of \mos. To remove any further adsorbates from the sample, we annealed it to 570\,K. Sulphur vacancies were then introduced by sputtering with Ne ions ($\sim 550$\,eV) at room temperature at 45\textsuperscript{$\circ$}-angle for a few seconds. After confirming the presence of an appropriate density of S vacancies by STM, \cfth\ molecules were evaporated from a Knudsen cell at an evaporation temperature of 365\,K onto the \mos/Au(111) sample cooled to 160\,K with subsequent annealing to 220\,K. For details on  the synthesis of the \cfth\ molecules, see Supplementary Information (SI), section VI.  All STM measurements were performed at 4.5\,K under UHV conditions and \dIdV spectra were recorded using a standard lock-in technique.

For computational analysis, we conducted an exhaustive \textit{ab initio} random structure search with the \textit{GenSec} package~\cite{maksimov_gensec, maksimov_thesis}. For molecular placement and distortion, random values were assigned to the internal torsional angles connecting the benzene rings of the molecule and to the orientation of the molecule with respect to the MoS$_2$ substrate. We constrained the search space by requiring that at least one atom of the molecule be placed at a distance of 3\,\AA~from the defect site. A $6 \times 6$ unit cell containing one sulfur vacancy was assumed and, in the first step, no doping of the substrate was considered. Geometry relaxations were performed only on structures that met the criteria of no steric clashes according to a database of scaled van der Waals radii. The initial relaxations were performed with the PBE functional and MBD-NL van der Waals corrections~\cite{hermann+prl2020}, \textit{light} settings in FHI-aims~\cite{blum+cpc2009} and a $1\times1\times1$ k-grid. We conducted searches with the 3P-S molecule, the \cfth\ molecule and its dehydrogenated derivative, the \cft\ molecule. We optimized 35 structures of 3P-S, 45 structures of \cfth\ and 75 structures of \cft\ in each search. 

To increase the conformational diversity, we added CF$_3$ groups and H atoms to transform 3P-S molecules into \cft\ and \cfth\ and post-relaxed them. More details about the structure search can be found in the SI, section~I. We ended up with one candidate structure of \cfth\ and five candidate structures of \cft\ on \mos\ that anchored on the defect.
This subset of structures was further optimized with \textit{tight} settings and a $4\times4\times1$ k-grid, and then post-optimized with the HSE06~\cite{krukau+jcp2006} functional with MBD-NL van der Waals corrections and \textit{intermediate} settings.   Electronic structure properties and further analysis were all conducted at the HSE06 level of theory, in addition to a post-processing spin-orbit coupling correction. 

The same set of 6 conformers was re-optimized including a negative charge in the substrate, to model a negatively charged vacancy with charge $-1$. This was achieved by the use of a virtual-crystal approximation (VCA) scheme, as explained in Ref.~\cite{Akkoush2023}. We carried out geometry optimizations with the HSE06 functional with MBD-NL van der Waals corrections and \textit{intermediate} settings, including the extra charge. Subsequently, electronic structure properties were analyzed at this level of theory, including the spin-orbit coupling correction. 

Binding energies of anchored molecular structures under ultra-high-vacuum conditions (partial pressure tending to zero) were calculated by
\begin{equation}
E_b = E_{\text{anchor}}(q) - E_{\text{MoS}_2+\text{VS}}(q) - E_{\text{mol}} + q E_f^{\text{Au}}
\end{equation}
where $q$ is the charge, $E_{\text{anchor}}$ is the total energy of the anchored molecular system, $E_{\text{MoS}_2+\text{VS}}$ is the total energy of the MoS$_2$ substrate containing the S vacancy, $E_{\text{mol}}$ is the total energy of the \cfth~ molecule in the gas-phase and $E_f^{\text{Au}}$ is the Fermi energy of Au. The total energies were obtained with the HSE06+MBD-NL functional and the Au Fermi energy at the HSE06 level of theory (-4.95\,eV) was considered as an internally consistent reference. The value entering the computation was referenced to the position of the valence-band-maximum (VBM) of the MoS$_2$ substrate containing the S vacancy at a consistent charge state and same DFT level of theory. With this definition, a more negative $E_b$ means stronger binding. When the anchored molecule was dehydrogenated as \cft, we considered an adsorbed hydrogen far away from the defect site to obtain a simple estimate of a lower bound of the binding energy. We did not correct for finite size effects.

\section{Anchoring \cfth\ molecules to S vacancies in \mos}
The pristine \mos\ islands appear in STM topography with a characteristic moiré pattern caused by the lattice mismatch between \mos\ and the underlying Au(111) \cite{Krane2018a} (Fig.\,\ref{fig:prep_overview}a). After sputtering with Ne ions, the \mos\ layer is decorated with defects appearing with a triangular rim around a dark center (see Fig.\,\ref{fig:prep_overview}b). These defects have been attributed to S vacancies in the terminating S layer~\cite{Trishin2023b}. Importantly, at low temperatures these sites were shown to localize a single electron, which is exchange coupled to the metal electrons at the interface of \mos\ and Au(111) as reflected by a Kondo resonance~\cite{Trishin2023b}. The localized negative charge is most likely transferred from the metal substrate. The observed charge state is in agreement with theoretical predictions, favoring the stabilization of the charged defect state in proximity to a Au substrate~\cite{Akkoush2023, tan+prm2020, tumino+jpcc2020}. 

Large organic molecules such as \cfth\ have a low diffusion barrier on \mos\ \cite{Krane2018}. To reduce diffusion and maintain a few molecules on the \mos\ islands, a deposition temperature of 160\,K was chosen. However, we then observed that most molecules were physisorbed on the islands with a very large mobility under the influence of the tip. Subsequent annealing to 220\,K led to major diffusion, so that most of the \cfth\ molecules were found on the Au(111) substrate (forming clusters) rather than on the \mos\ islands as revealed in Fig.\,\ref{fig:prep_overview}c. Yet, we also detected a few oval-shaped protrusions indicating single molecules on the \mos\ substrate. 

\begin{figure}
  \centering
\includegraphics[width=0.95\linewidth]{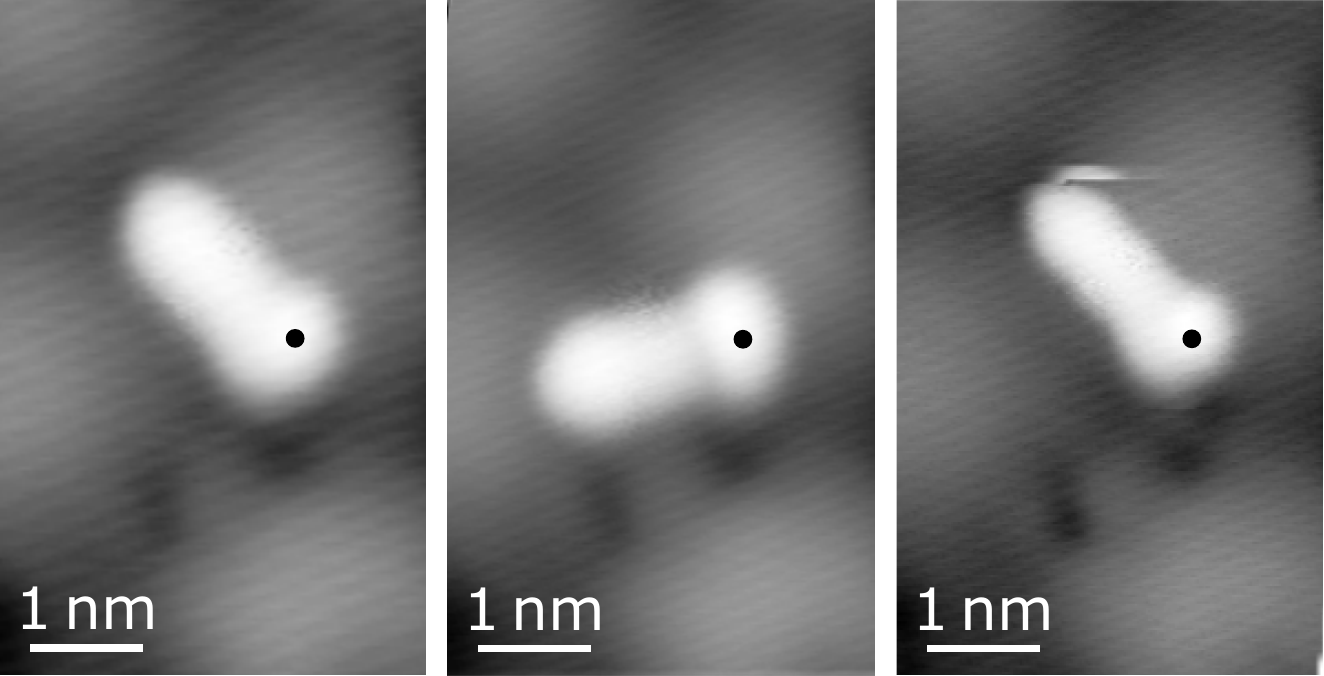}
\caption{STM topographies of subsequent rotation steps of a \cft\ molecule on \mos/Au(111) around its anchoring point (marked by black dot) after interaction with the STM tip
  (setpoints: (a) and (c) 1\,V, 40\,pA, (b) 1\,V, 50\,pA).}
\label{fig:rotation}
\end{figure}

A close-up view on one of these molecules is shown in Fig.\,\ref{fig:rotation}. Interestingly, the molecule is free to rotate around one of its terminations (indicated by a black dot in the series of images in Fig.\,\ref{fig:rotation}). The rotations can be induced by approaching the STM tip.  In none of the cases did the molecule move laterally. We thus assume that it is anchored to a defect site. Most probably, the defect is a S vacancy in the top layer of \mos, as we did not observe anchored \cfth\ molecules without prior creation of these sites by Ne bombardment. The robust anchoring of the molecule suggests the presence of a covalent bond to the S-defect.
Our preparation procedure of deposition at low temperature followed by mild annealing thus allows for some of the \cfth\ to bond to the S vacancies.

To unravel the atomic structure arising from this chemisorption scenario, we performed a combination of first-principles simulations. We analyzed the outcome of structure searches performed on a neutral and a negatively charged S-vacancy site in a free-standing monolayer of \mos\ in contact with intact \cfth\ molecules and dehydrogenated \cft\ molecules. We considered both scenarios because in a previous publication from some of the authors~\cite{Akkoush2023}, we have established that the formation energy difference between neutral and negatively-charged vacancies only amounts to 43\,meV when considering the Au substrate as the electronic reservoir. Therefore, by estimating the respective Boltzmann weights, we conclude that while at very low temperatures one would  exclusively observe charged vacancy sites, at the reaction temperature of 220\,K, around 10\% of these vacancy sites would be neutral.

When considering a negative charge, we did not obtain any \cfth\ molecules docked to the vacancy. In contrast, we obtained stable dehydrogenated \cft\ molecular configurations anchored to the defect site. 
This suggests that if the molecule docks on a negatively-charged vacancy, the H atom can be dissociated from the thiol group with the dangling bond at the S termination then being the ideal candidate for healing the S vacancy site in \mos.
In the case of a neutral vacancy, both \cft\ and \cfth\ conformers are stable (see SI, Section I). 

Regarding the anchoring mechanism, we note that in order for \cft\ to bond to the vacancy site, a reaction barrier for dehydrogenation has to be overcome. We refer to the work of Li \textit{et al.} who characterized viable pathways for this reaction to happen at the defect site where the thiol group  binds~\cite{LiAngew2017}, reporting barriers of 0.2-0.3\,eV. We propose that these barriers can be lowered at negatively-charged defects because these sites will be more attractive to protons. The abstracted H is likely to be highly mobile on \mos\ and most probably diffuses to the metallic substrate. We surmise that a reaction, where H stays at a defect site and the unstable neutral radical diffuses and eventually finds another free defect site, is energetically highly unlikely. Indeed, negatively-charged thiolates have been reported to be repelled even by neutral defects~\cite{LiAngew2017}. 

The local charge of the hybrid state formed by the molecule anchored to the defect site in \mos\ is independent on whether the initial vacancy defect was neutral or negatively charged. Instead, it will be dictated by the stability of the hybrid state with respect to the electronic reservoir. Taking the electronic reservoir to be the Au substrate, we obtain the binding energy of the most favorable neutral conformer (considering \cfth\, and \cft) to be -0.56\,eV, while the most favorable negatively charged case (considering only \cft, as the anchored \cfth\ is not stable in a negatively charged state) leads to a binding energy of only -0.08\,eV, at the verge of stability within the uncertainties of our estimation. Note that the latter value does not include the energy penalty for dehydrogenation of \cfth\ into \cft. We did not explore positively charged final states because they would be even less likely in this setting. We thus propose that the final state of the anchored molecules is neutral. %The negatively charged anchored molecules and the respective electronic states we considered are shown in the SI, section V.

We now discuss the structural conformations of the anchored molecules. The most stable \cfth\ conformer presents a tilt angle of 15$^\circ$ with respect to the surface normal. Regarding the dehydrogenated \cft\ species, we found upright-standing configurations to be energetically favored. However, other stable minima within a 300\,meV energy window are also found at different degrees of inclination toward the surface~(see SI). In general, conformations with larger tilting angles can be stabilized when considering entropic contributions. This was confirmed by exploratory molecular-dynamics simulations at 100-300\,K, presented in the SI, section II. One can easily picture this, as any tilted configuration has access to a larger configurational space of rotations around the normal axis. Of course, we also cannot exclude that considering a lower concentration of molecules in the simulations can decrease the enthalpic penalty of tilted configurations. 

In summary, we propose that, for either \cft\ or \cfth\ anchored to defect sites, three factors contribute to the stabilization of the tilted conformations which are observed experimentally: (i) a low concentration of molecules, (ii) entropic contributions that thermally populate the softer rovibrational modes of the molecule, (iii) lowering of zero-point-energy contributions when molecular groups are physisorbed on the surface.

\begin{figure}
  \centering
  \includegraphics[width=0.95\linewidth]{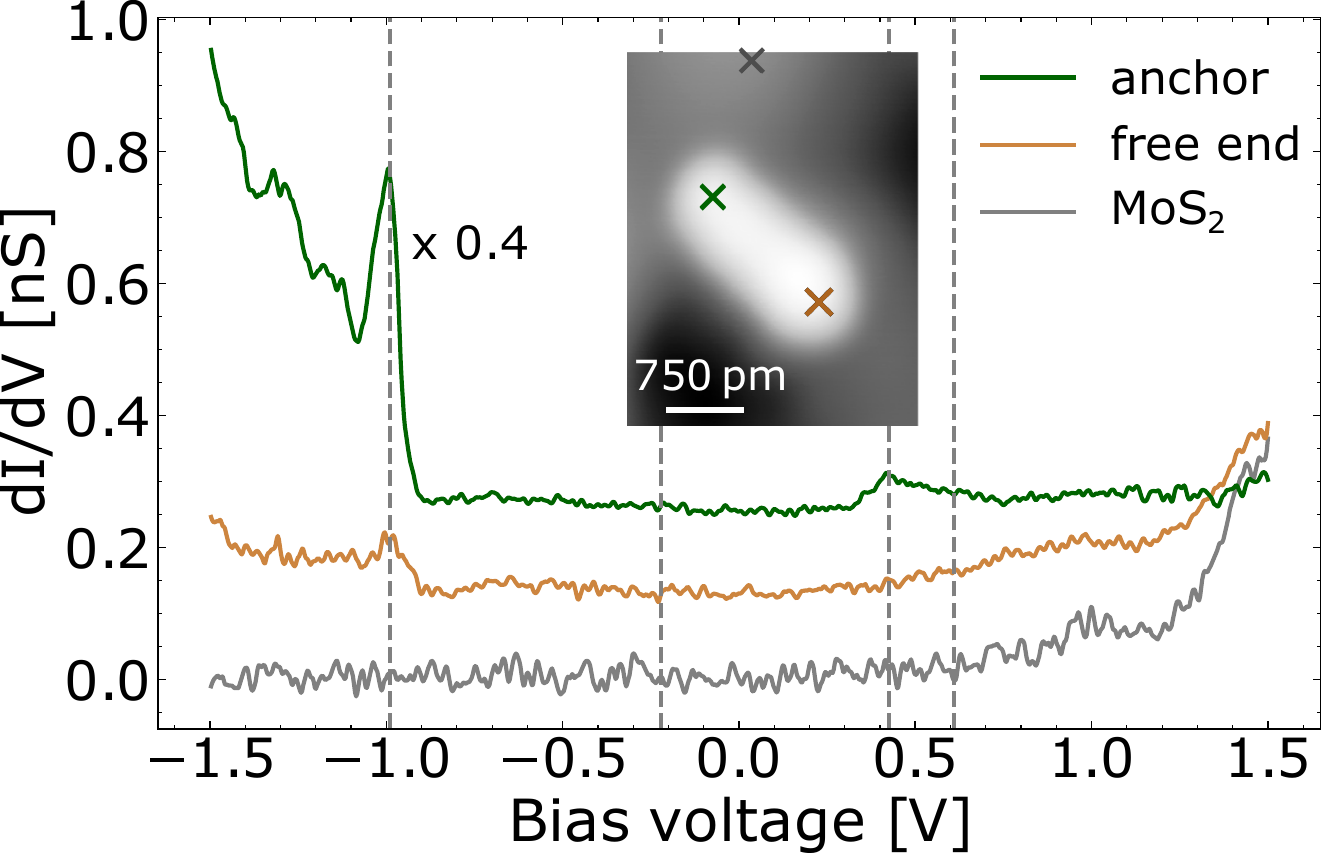}
  \caption{  \dIdV\ spectra on an anchored \cft\ molecule on \mos/Au(111), on the anchor point (green), the free end (orange) and the reference taken on bare \mos\ (grey). The spectra are vertically offset by 0.125\,nS from each other. The spectrum on the anchored side has been rescaled by a factor of 0.4 for better comparison with the other spectra. The grey dashed lines denote the bias values of the \dIdV\ maps taken in Fig.\,\ref{fig:maps}.
  (Feedback opened at 1.5\,V, 100\,pA, lock-in parameters: 2\,mV (\mos\ and anchor) and 5\,mV (free end) modulation amplitude.)
  Inset: STM topography of the measured molecule, with colored crosses marking the positions where the spectra are recorded (setpoint: 2\,V, 30\,pA).%\\
  }
\label{fig:spec}
\end{figure}

\section{Electronic properties of anchored molecules}
To investigate the electronic properties of the anchored molecules, we record \dIdV\ spectra on the anchored molecules. 
As we will show below, we found two types of \dIdV spectra that we associate to two distinct molecules anchored to the S defects according to our theoretical simulations. One type consists of an intact \cfth\ molecule, whereas the other type has undergone dehydrogenation to a  \cft\ molecule.

\subsubsection{Anchored \cfth\ molecules}
We start by presenting the \dIdV spectra on the more abundant species. Spectra on the anchoring site and the center of the molecule are shown in Fig.\,\ref{fig:spec} (green and orange) together with a spectrum on the pristine \mos\ (grey). In the latter, the conductance is essentially flat over a large bias-voltage range, in agreement with the semiconducting bandgap of \mos, with the onset of the conduction band smeared out while exhibiting a small peak at $\sim$\,0.9\,eV, followed by a stronger one at $\sim$\,1.4\,V~\cite{Krane2018}. 

\begin{figure*}
  \centering
\includegraphics[width=\linewidth]{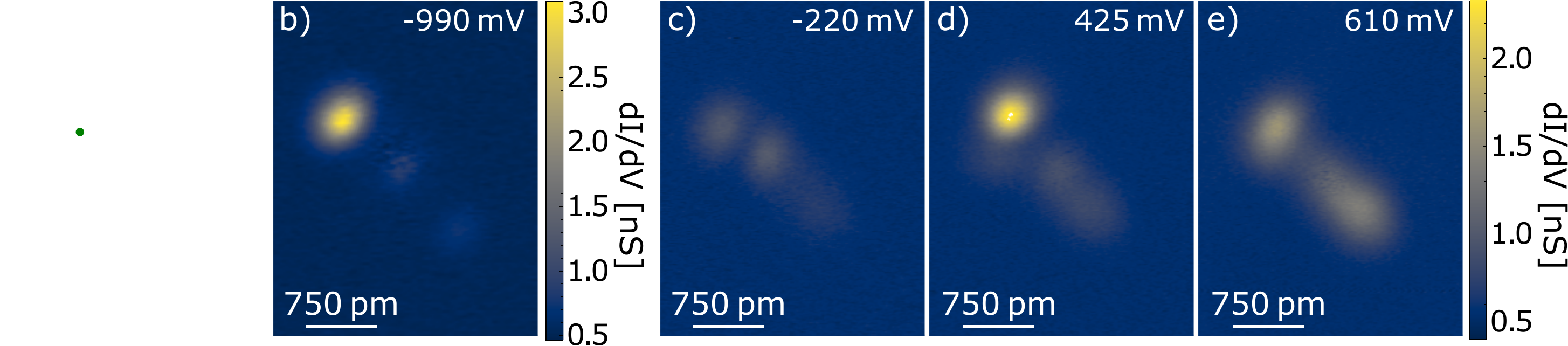}
\caption{a) STM topography (setpoint: 2\,V, 30\,pA) and constant height \dIdV\ maps of an anchored \cft\ molecule (anchorpoint marked by green dot) taken at energies marked by grey lines in Fig.\,\ref{fig:spec}: b)~-990\,mV, c) -220\,mV, d)~425\,mV, e)~610\,mV (Feedback opened at (b) -990\,mV, 30\,pA over anchored end, (c)-(e) 610\,mV, 30\,pA above free end, lock-in parameters for all: 5\,mV modulation amplitude).}
\label{fig:maps}
\end{figure*}

The anchored molecules show two main features inside the bandgap of \mos\ (Fig.\,\ref{fig:spec}): A positive ion resonance (PIR) of high intensity at $\sim-1$\,V with satellite peaks following behind, and a broader negative ion resonance (NIR) starting at $\sim 420$\,mV. Both states are found to be very intense at the anchored end (green spectrum in Fig.\,\ref{fig:spec}), and show nearly no intensity on the free end of the molecule (orange spectrum in Fig.\,\ref{fig:spec}). To bring out the spatial distribution of these states more clearly, we present constant-height \dIdV\ maps recorded at the resonance energies in Fig.\,\ref{fig:maps} (energies of maps marked with a grey dashed line in Fig.\,\ref{fig:spec}).

\begin{figure}
  \centering
  \includegraphics[width=\columnwidth]{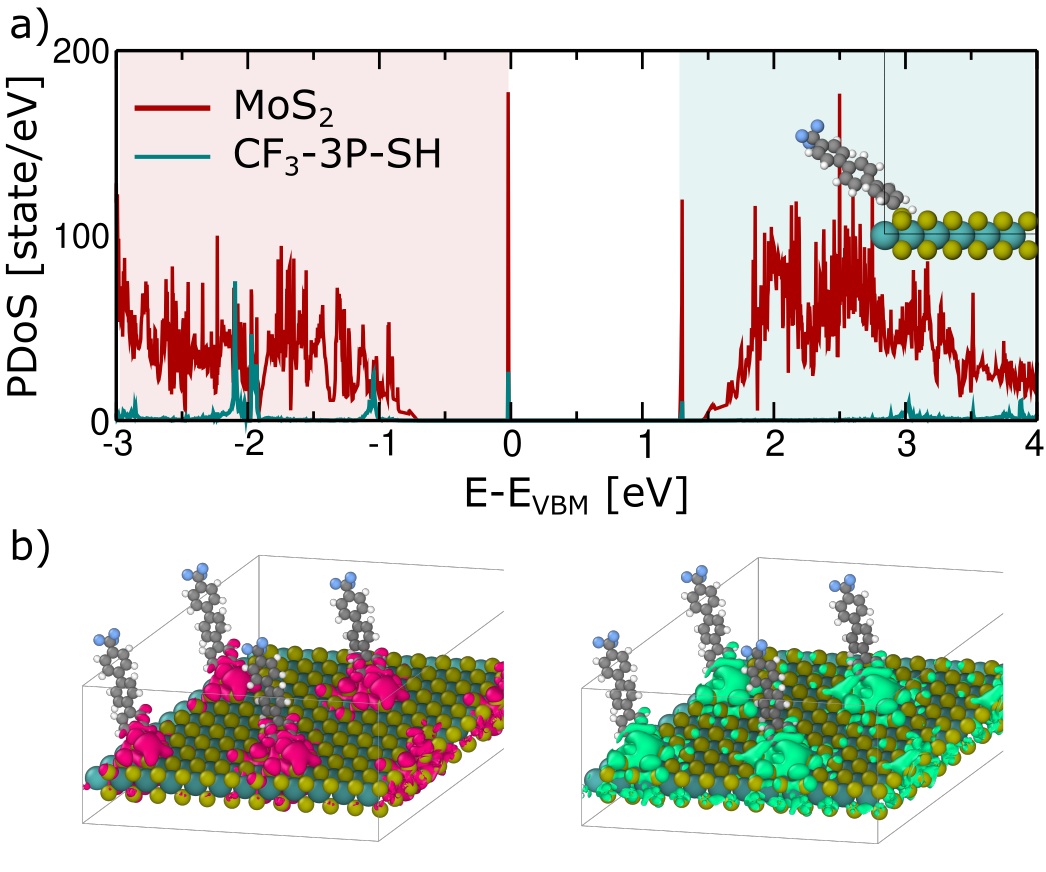}
  \caption{a) Projected electronic density of states (PDOS) of the stable conformer of \cfth~anchored to the neutral \mos~monolayer, obtained with density-functional theory and the HSE06 exchange-correlation functional. The molecular geometry is shown in the inset. Cyan and red lines represent the PDOS on the molecule and on \mos\, respectively. Red shaded area marks occupied states and blue shaded area marks unoccupied states. b) Orbital densities corresponding to the highest doubly-occupied state in pink and the lowest unoccupied state in green.  We show an isodensity of 0.0002 e/Bohr$^3$.  %\\
  }
\label{fig:theory-neutral}
\end{figure}

The map taken at the energy of the PIR (-990\,mV, Fig.\,\ref{fig:maps}b) shows a strong localization at the anchoring site. Additionally, it shows some faint intensity along the molecular backbone, exhibiting two nodal planes. The NIR at $\sim 420$\,mV also shows the largest contribution close to the anchoring site, but slightly offset from the molecular axis (Fig.\,\ref{fig:maps}d). Maps at low energy (-220\,meV, Fig.\,\ref{fig:maps}c) and higher energy (610\,meV, Fig.\,\ref{fig:maps}e) show a faint structure along the molecule with some intramolecular modulation.

\begin{figure}
  \centering
\includegraphics[width=.95\linewidth]{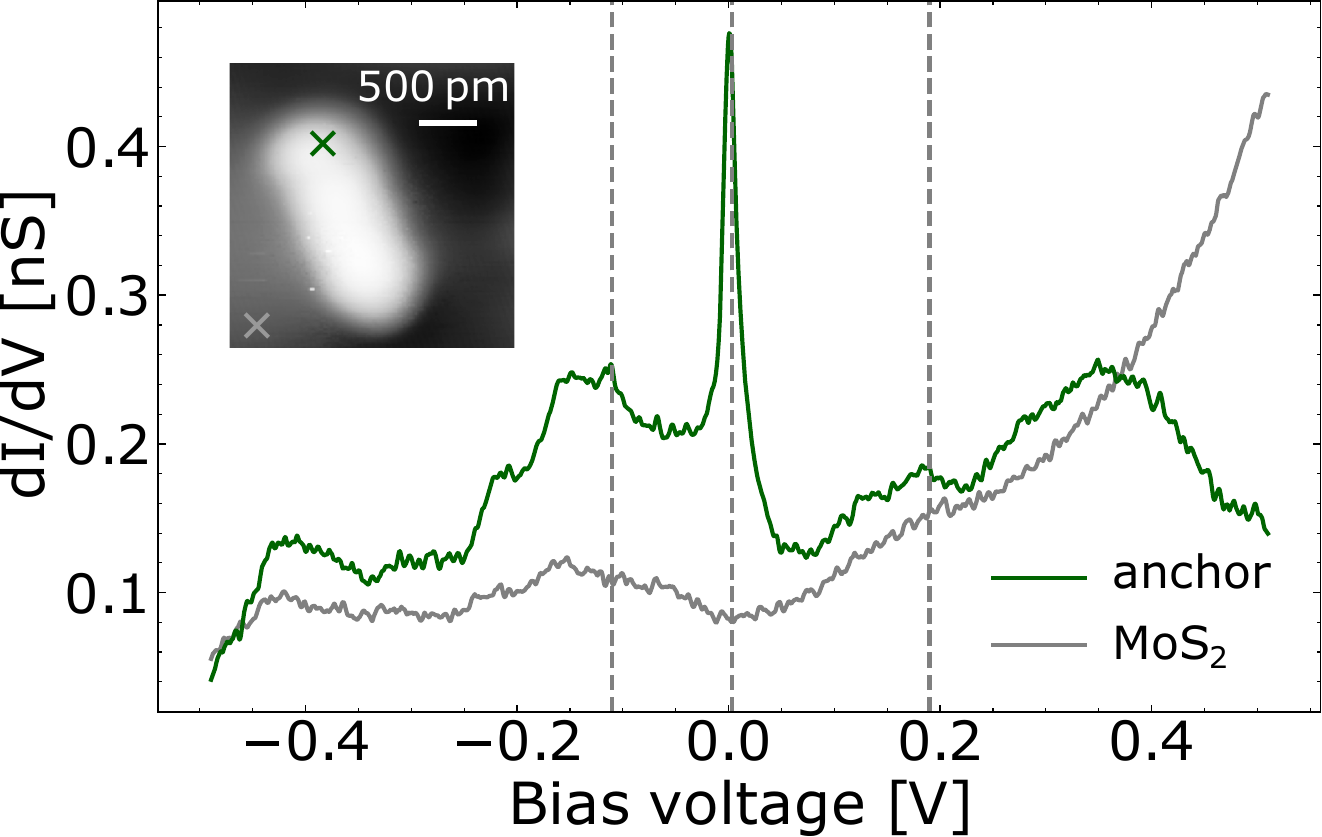}
\caption{\dIdV\ spectra on an anchored molecule on \mos/Au(111), on the anchor point (green) and the reference taken on bare \mos\ (grey). The grey dashed lines denote the bias values of the \dIdV\ maps shown in Fig.\,\ref{fig:maps_Kondo}.
  (Feedback opened at 500\,mV, 100\,pA, lock-in parameters: 5\,mV modulation amplitude.)
  Inset: STM topography of the measured molecule, with colored crosses marking the positions where the spectra were taken (setpoint: 1.3\,V, 30\,pA).}
\label{fig:spec_Kondo}
\end{figure}

\begin{figure*}
  \centering
\includegraphics[width=\linewidth]{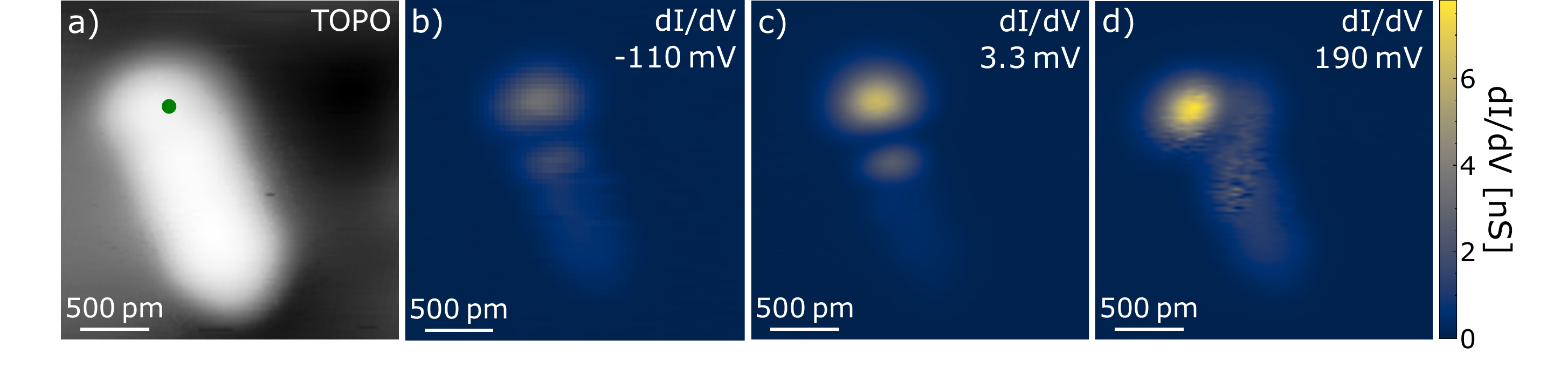}
\caption{a) STM topography (setpoint: 1.3\,V, 30\,pA) and constant-height \dIdV\ maps of an anchored molecule (anchor point marked by green dot) taken at energies marked by grey lines in Fig.\,\ref{fig:spec_Kondo}: b) -110\,mV, c) 3.3\,mV and d) 190\,mV (feedback opened at 500\,mV, 100\,pA and 100\,pm tip approach, lock-in parameters: 5\,mV modulation amplitude).}
\label{fig:maps_Kondo}
\end{figure*}

\begin{figure}
  \centering
  \includegraphics[width=\columnwidth]{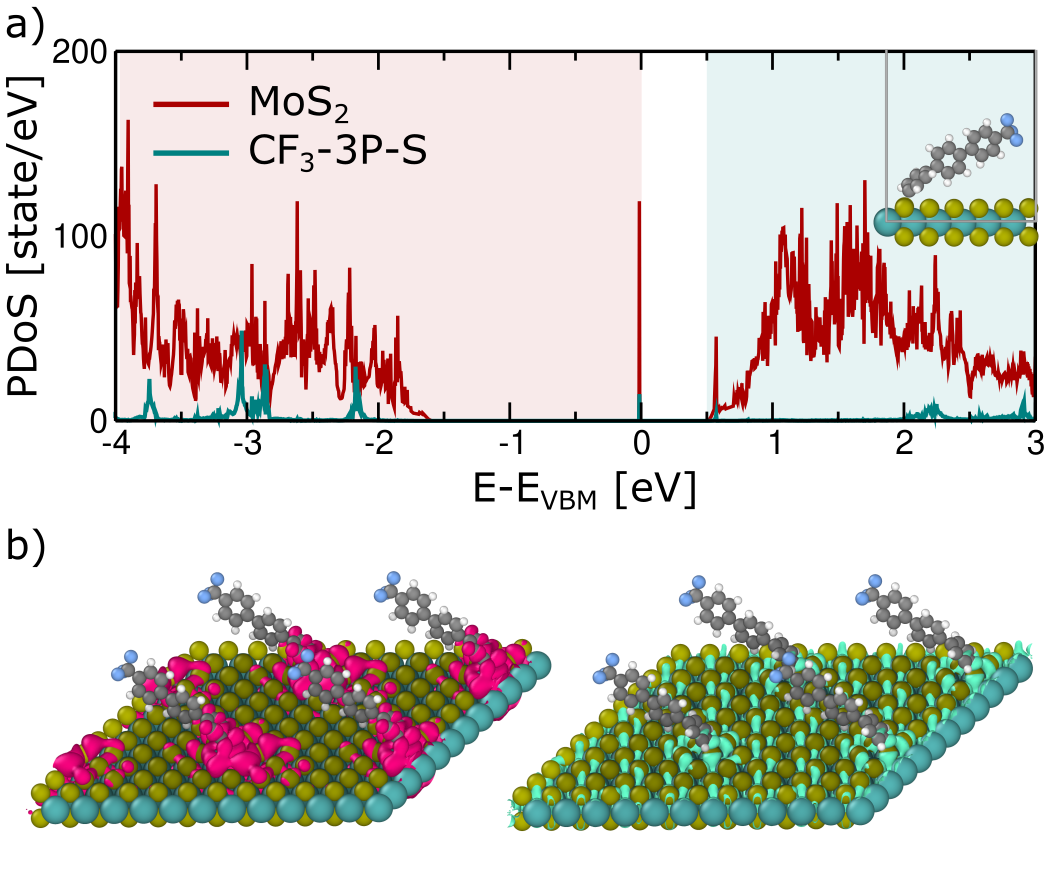}
  \caption{a) Projected electronic density of states (PDOS) of the stable conformer of \cft~anchored to the neutral \mos~monolayer, obtained with density-functional theory and the HSE06 exchange-correlation functional. The molecular geometry is shown in the inset. Cyan and red lines represent the PDOS on the molecule and on \mos\, respectively. Red shaded area marks occupied states and blue shaded area marks unoccupied states. The state at 0\,eV is singly occupied. b) Orbital densities corresponding to the highest singly-occupied state in pink and the lowest fully unnocupied state at 0.57\,eV in green.  We show an isodensity of 0.0002\,e/Bohr$^3$.  %\\
  } \label{fig:kondo-theory}
\end{figure}

To understand the origin of the observed states, we analyze the electronic structure obtained by DFT calculations of candidate molecular configurations anchored on the S vacancy.
We first consider the anchoring of an intact \cfth\ molecule on a neutral vacancy. This choice is motivated by the absence of a Kondo resonance, and, hence, of a singly occupied state.

In Fig.\,\ref{fig:theory-neutral} we show the corresponding density of states of the \cfth\ molecule adsorbed on the vacancy.  
As we neglect the Au(111) substrate in the simulations, we do not expect quantitative agreement in the energy-level alignment but aim for a qualitative identification of the hybrid electronic states. In this case, both the highest (doubly-)occupied state and the lowest unoccupied state are localized at the anchoring site. These states are separated by an energy gap of 1.3\,eV. The states originate from the hybridization of the frontier orbitals of the \cfth molecule. The energy gap between PIR and NIR and the localisation thus agree well with the observations in experiment. We recall that for \cfth\ to bond to the defect site, the vacancy must be neutral. Although all pristine S vacancies exhibit a Kondo resonance at low temperatures reflecting a charged state, at the annealing temperature of 220\,K the vacancy is transiently neutral with a relevant probability of 10\%, as discussed above. The neutral state can thus act as an anchoring site for the intact \cfth\ molecule.

\subsubsection{Anchored dehydrogenated \cft\ molecules}

In contrast to the molecular species described above, we find a few molecules with a very different spectral fingerprint. As shown in Fig.\,\ref{fig:spec_Kondo}, these molecules exhibit a sharp resonance at zero bias that we associate to a Kondo resonance flanked by two side peaks at $\sim -110$\,mV and $\sim 190$\,mV. Differential conductance maps at these energies are presented in Fig.\,\ref{fig:maps_Kondo} along with the topographic image. All resonances are mainly localized at the anchoring site. Given their similar spatial distribution and close energy spacing, we suggest that the side peaks arise from tunneling through the singly occupied state at negative bias and the doubly occupied state separated by the Coulomb charging energy from the singly-occupied state at positive bias. The experiments thus indicate the presence of a singly-occupied state when the molecule is anchored to the S vacancy. 

Following our stability analysis above, the hybrid state must be charge neutral, but a singly-occupied state must be present. Therefore, we consider a dehydrogenated \cft\ radical bonded to the S vacancy. 
We suggest that H is catalytically dissociated from the thiol group when the molecule is in the vicinity of a negatively charged vacancy. The dangling bond is then saturated upon chemisorption of the \cft\ molecule to the vacancy site, and H could subsequently diffuse away from the anchored molecule. Because this incurs an energy penalty, this anchoring mode is less likely, in agreement with the less frequent observation as opposed to the first species discussed above. Overall charge neutrality results from the interaction of the Au substrate acting as a charge reservoir as discussed above. 
A representative molecular geometry, the projected electronic density of states and the visualization of key electronic states are shown in Fig.~\ref{fig:kondo-theory}. 
The bonded conformations show a singly-occupied state in the gap that could lead to a Kondo resonance. In agreement with experiment, this state is localized at the anchoring point. We also note that it is robust to several different molecular orientations found in the calculations. Other geometries with their corresponding electronic structure are reported in the SI, section III.

\section{Conclusions}
Using scanning tunneling spectroscopy, we have observed two types of reaction products of \cfth\ molecules anchoring to purposefully created  S vacancies in \mos\ on Au(111). Based on \textit{ab initio} structure searches, DFT calculations and comparison to the experimental tunneling spectra, we assign these two configurations to intact \cfth\ molecules or dehydrogenated \cft\ molecules and characterize their structural and electronic properties. 

Due to the presence of the Au(111) substrate acting as a charge reservoir, both locally-formed molecular hybrid states are charge neutral. Both species exhibit states within the bandgap of \mos and the anchored \cfth\ is more abundant. Interestingly, the \cft\ molecule features a singly-occupied state, which explains the few anchored molecules that present a Kondo resonance in experiment. We note that the precise energy alignment and the localization of these new hybrid states depends sensitively on the adsorption configuration of the molecule on the moir\'e structure of \mos/Au(111).

The combination of well-defined experiments and theory in this model system allowed an atomic-scale understanding of the properties of defect-engineered states, where a minimally biased first-principles structure search proved essential to unravel the possible atomic and electronic structure of docked molecules. Such an understanding would not be achievable with experiment or theory alone. 

Going forward, the use of molecules with functional end groups, such as charge donating or charge withdrawing groups, will enable further modulation of the energy level alignment. Attachment of molecular switches may additionally allow a reversible variation of energy levels and, thus, conductance properties, while magnetic molecular centers may merge magnetic with the optoelectronic properties of the TMDC layer for more complex device applications.

\section{Acknowledgements}
The authors thank Caterina Cocchi and Ana Valencia for discussions at an initial stage of this work, and Idan Tamir for technical support. We acknowledge funding by the Deutsche Forschungsgemeinschaft (DFG) through SFB 951 “Hybrid Inorganic/Organic Systems for Opto-Electronics” (project number 182087777, projects A13, A14, Z1).
\bibliography{bibliography}

\end{document}

\title{Supplementary Information: An atomic-scale perspective on individual thiol-terminated molecules anchored to single S~vacancies in \mos}%

\author{J.\,Rika Simon}
\affiliation{\mbox{Fachbereich Physik, Freie Universit\"at Berlin, 14195 Berlin, Germany}}

\author{Dmitrii Maksimov}
\affiliation{Max Planck Institute for the Structure and Dynamics of Matter, 22761 Hamburg, Germany}

\author{Christian Lotze}
\affiliation{\mbox{Fachbereich Physik, Freie Universit\"at Berlin, 14195 Berlin, Germany}}

\author{Paul Wiechers}
\affiliation{\mbox{Fachbereich Physik, Freie Universit\"at Berlin, 14195 Berlin, Germany}}

\author{Juan~Pablo~Guerrero~Felipe}
\affiliation{\mbox{Fachbereich Physik, Freie Universit\"at Berlin, 14195 Berlin, Germany}}

\author{Björn Kobin}
\affiliation{Institut für Chemie, Humboldt-Universität zu Berlin, 12489 Berlin, Germany}

\author{Jutta Schwarz}
\affiliation{Institut für Chemie, Humboldt-Universität zu Berlin, 12489 Berlin, Germany}

\author{Stefan Hecht}
\affiliation{Institut für Chemie, Humboldt-Universität zu Berlin, 12489 Berlin, Germany}
\affiliation{Center for Science of Materials, 12489 Berlin, Germany}

\author{Katharina J. Franke}
\affiliation{\mbox{Fachbereich Physik, Freie Universit\"at Berlin, 14195 Berlin, Germany}}
\email{franke@physik.fu-berlin.de}

\author{Mariana Rossi}
\affiliation{Max Planck Institute for the Structure and Dynamics of Matter, 22761 Hamburg, Germany}
\email{mariana.rossi@mpsd.mpg.de}

\maketitle

\section{\textit{Ab initio} random structure search}

\begin{figure}[h]
\centering
\includegraphics[width=0.9\textwidth]{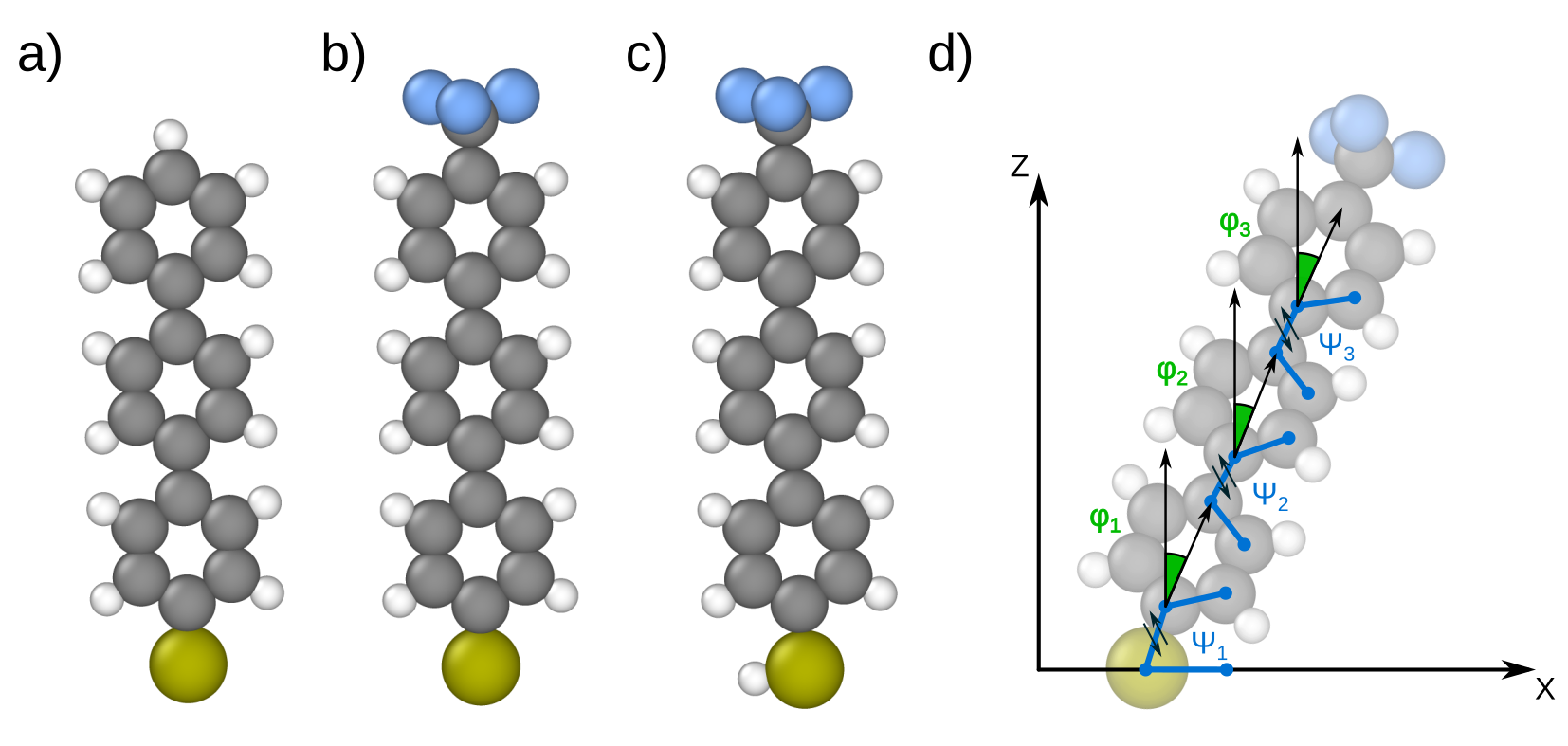}
\caption{Molecular structures of a) 3P-S, b) \cft\ and c) \cfth; d) notation of the angles and dihedral angles used in the work: $\varphi$ is the angle between the Z axis and the C-C vector formed by the carbon atoms connected to the other functional groups in the molecule, and $\Psi$ is the dihedral angle formed by four atoms either of the adjacent rings, or two atoms of the ring that is closest to the surface and two sulfur atoms from the surface. Colors of atoms: carbon is grey, sulfur is yellow, fluorine is blue and hydrogen is white.}
\label{SI:molecules_examples}
\end{figure}

Structure searches were conducted with the \textit{GenSec} package for three systems: 3P-S, \cft\ and \cfth\ in contact with a MoS$_2$ monolayer containing one neutral S vacancy. 
The three molecular structures are presented in Fig.\,\ref{SI:molecules_examples}. Different conformations of the molecules were obtained by randomly varying dihedral angles $\Psi_2$ and $\Psi_3$ (for definition see Fig.\,\ref{SI:molecules_examples}d). The molecules were then allowed to have a random orientation and position with respect to the MoS$_2$ surface with the constraint that at least one atom of the molecule was not further than 3\,\AA~away from the S vacancy. Further details are found in the main text.

\begin{figure}[ht]\centering
\subfloat[]{\label{SI:3P-S_energies}\includegraphics[width=.32\linewidth]{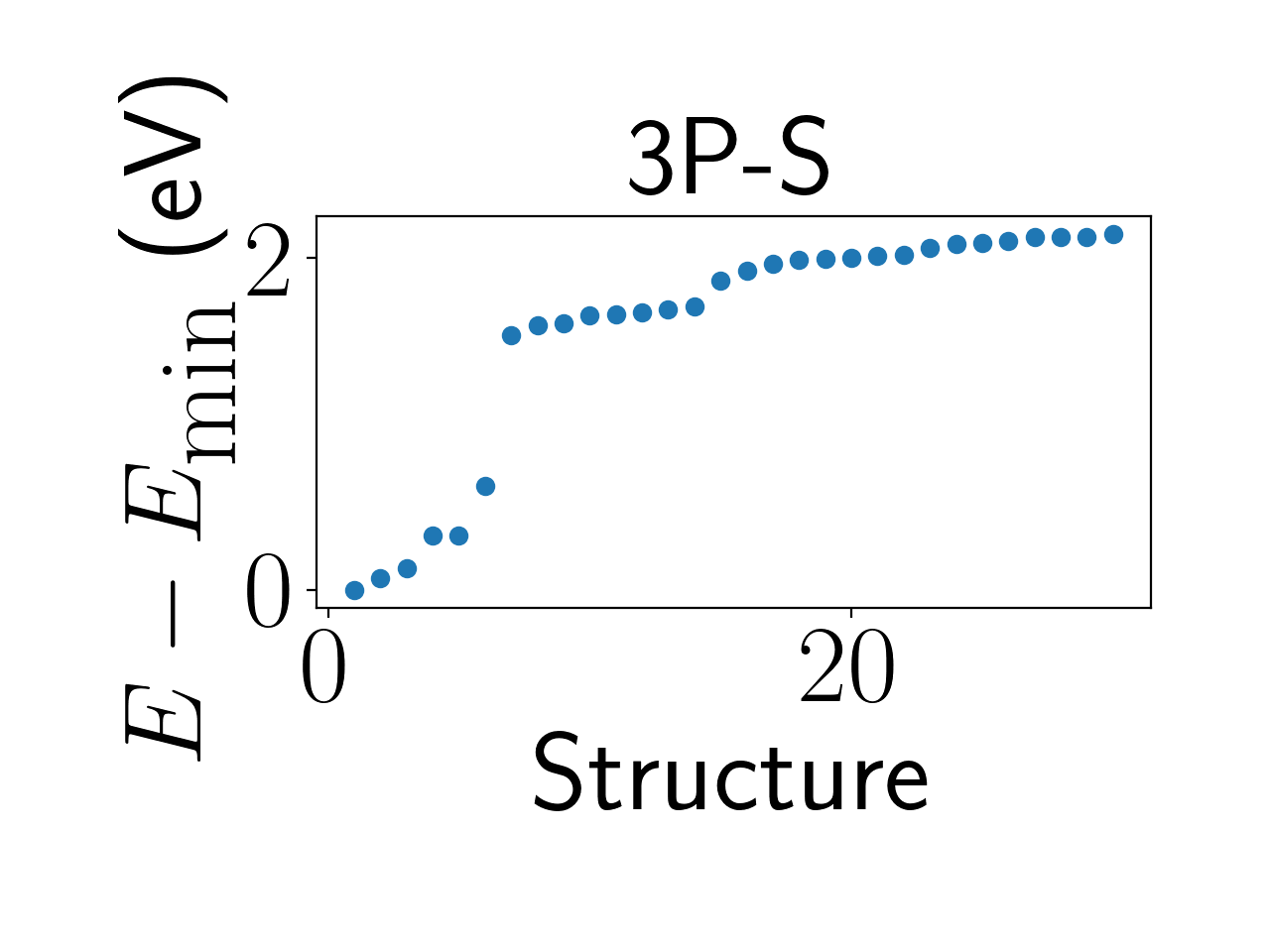}}\hfill
\subfloat[]{\label{SI:CF3-3P-S_energies}\includegraphics[width=.32\linewidth]{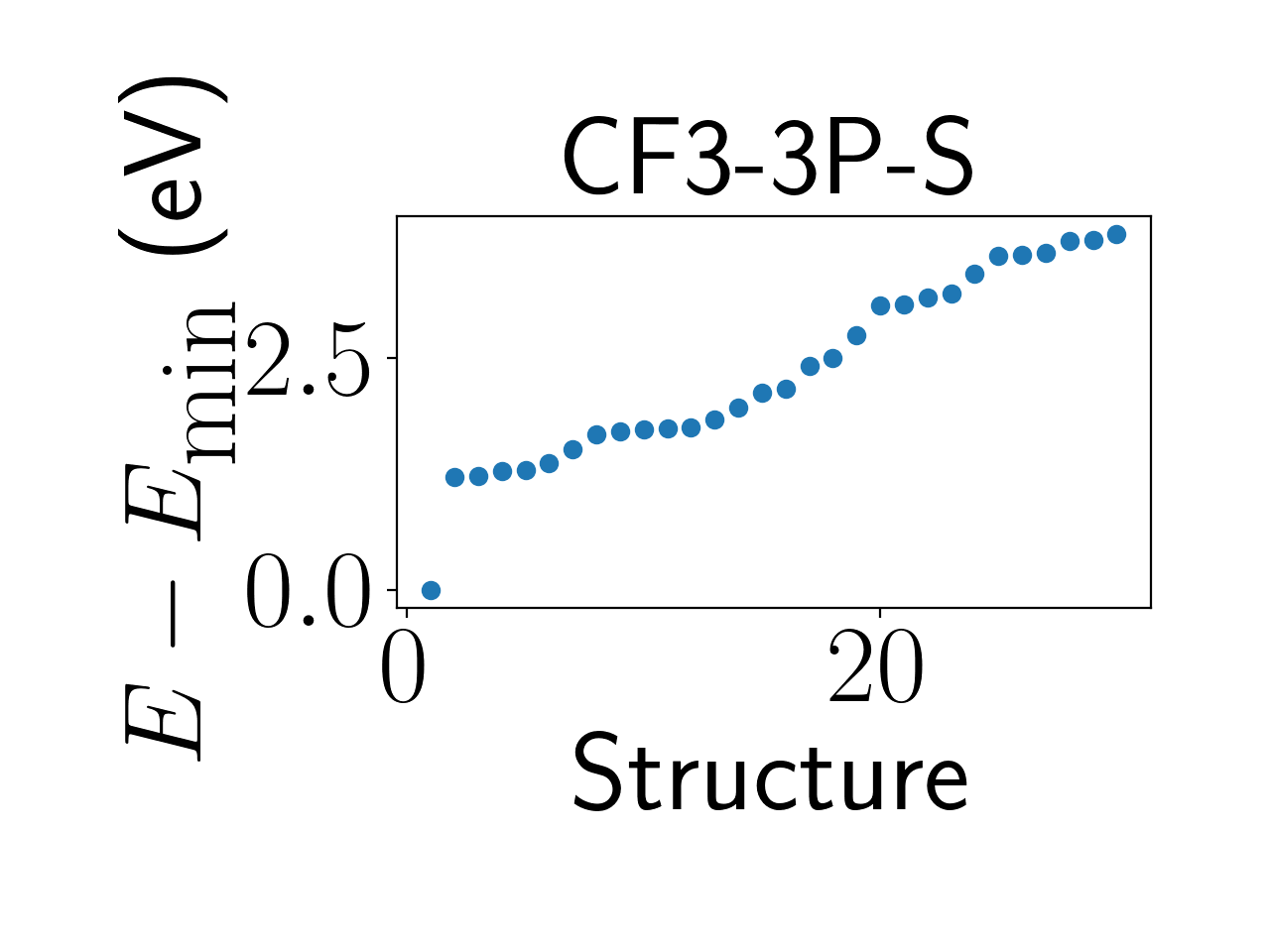}}\hfill 
\subfloat[]{\label{SI:CF3-3P-SH_energies}\includegraphics[width=.32\linewidth]{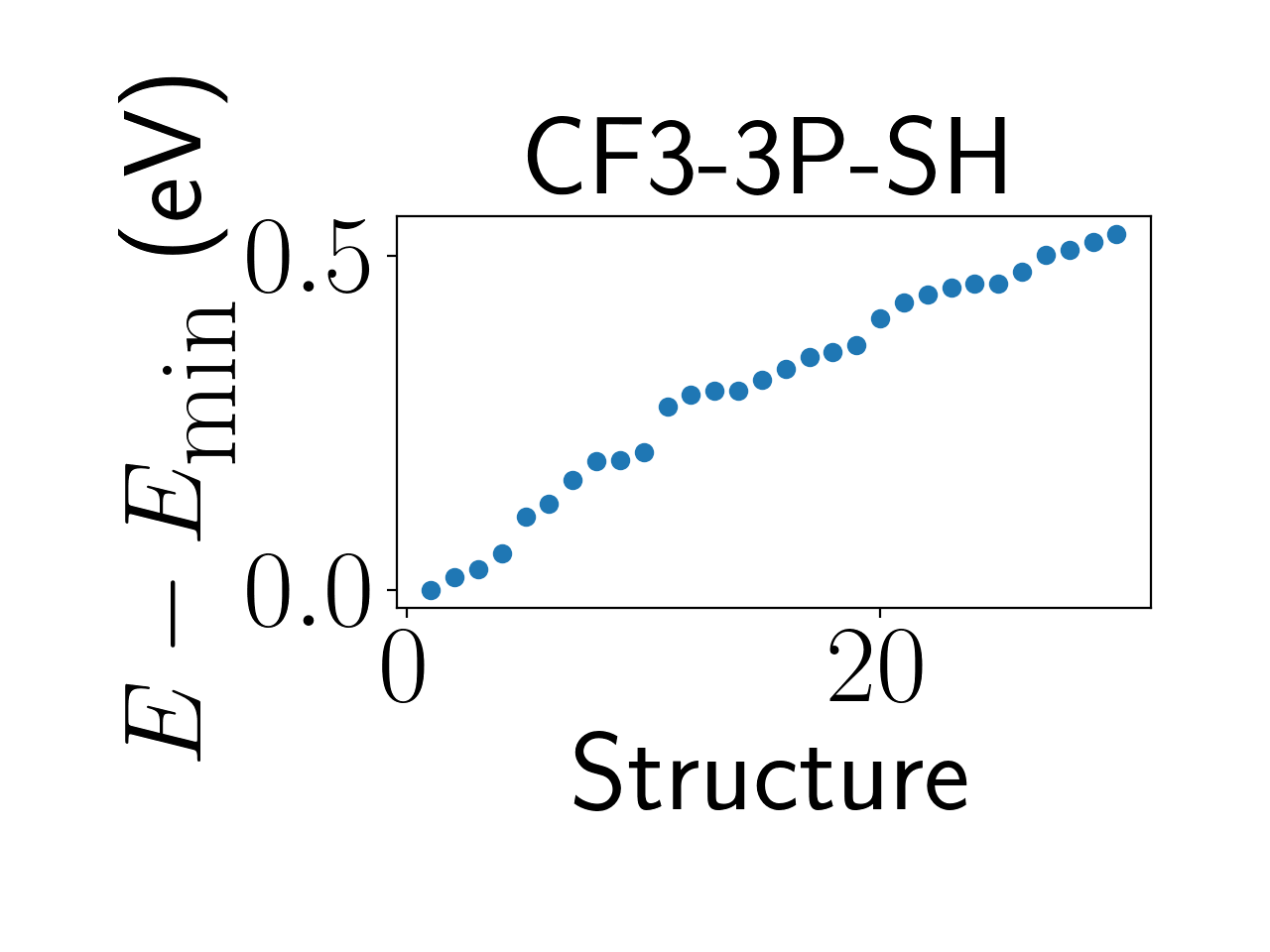}}
\caption{Relative energies of 3P-S, \cft\ and \cfth\ conformers in contact with the MoS$_2$ substrate with one neutral S vacancy, as obtained after the random structure search.} 
\label{SI:Energies}
\end{figure}

\begin{figure}[ht]
\centering
\includegraphics[width=0.95\textwidth]{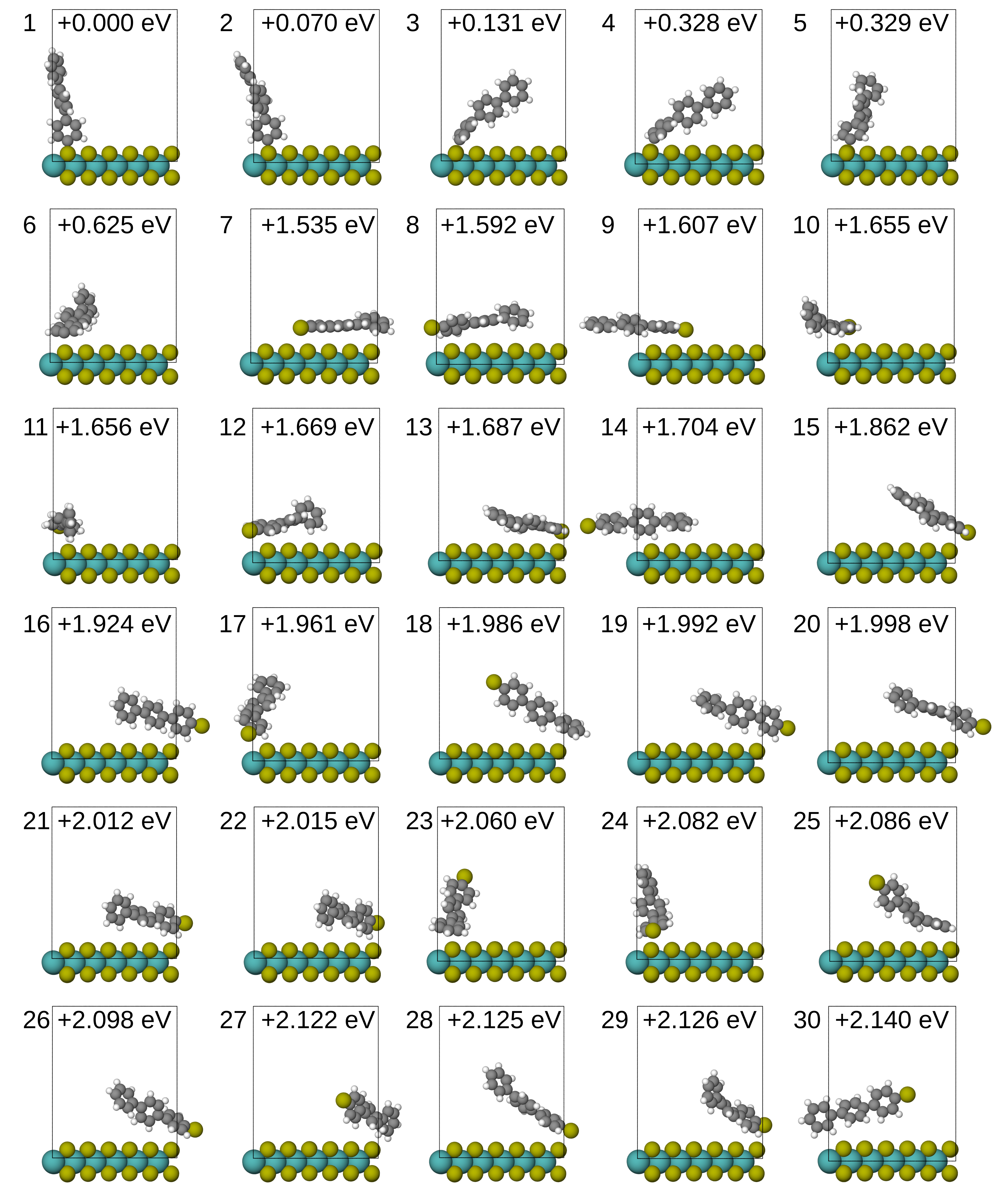}
\caption{Structures of 3P-S adsorbed or physisorbed on the MoS$_2$ surface containing an S vacancy. Colors of atoms: carbon is grey, sulfur is yellow, molybdenum is cyan, and hydrogen is white.}
\label{SI:3P-S_sides}
\end{figure}

\begin{figure}[ht]
\centering
\includegraphics[width=0.95\textwidth]{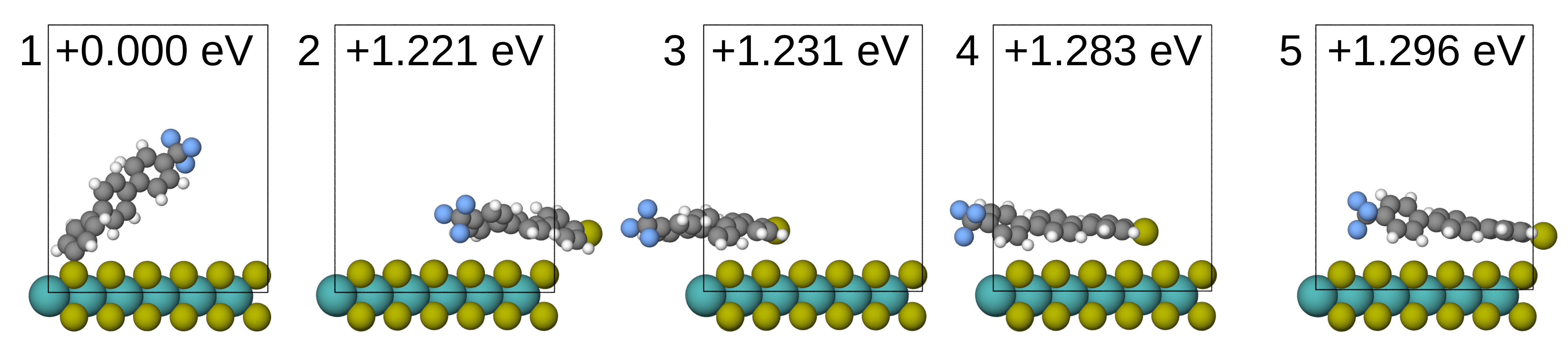}
\caption{Structures of \cft\ adsorbed or physisorbed on the MoS$_2$ surface containing the S vacancy. Color code like in other figures. These are the five lowest energy candidates.}
\label{SI:CF3-3P-S_sides}
\end{figure}

\begin{figure}[ht]
\centering
\includegraphics[width=0.9\textwidth]{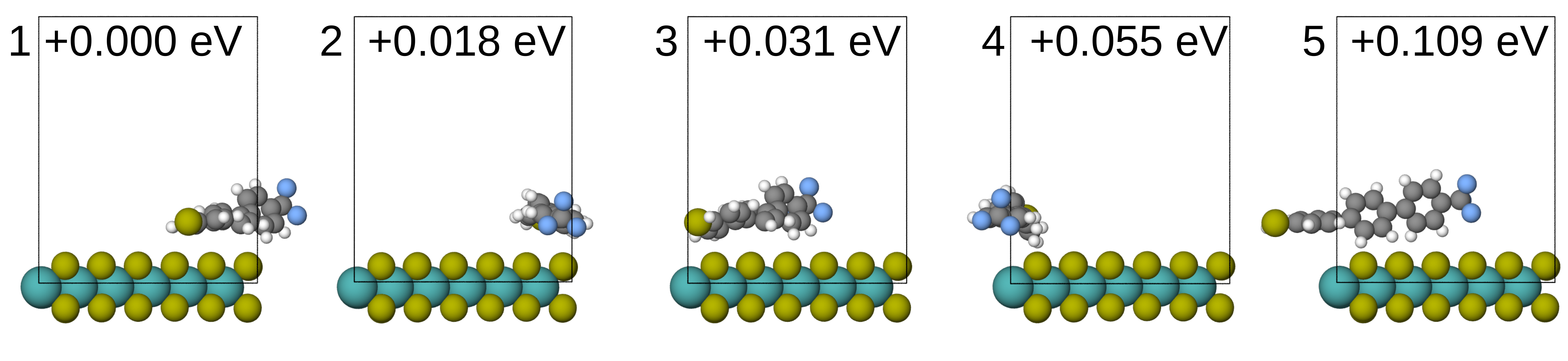}
\caption{Structures of \cfth\ physisorbed on MoS$_2$ surface containing the S vacancy. Color code like in other figures. These are the five lowest energy candidates.}
\label{SI:CF3-3P-SH_sides}
\end{figure}

Energy hierarchies obtained from the structure searches are presented in Fig.\,\ref{SI:Energies}. We obtained five distinct docked conformers for 3P-S and one docked molecule for \cft\ (none were found for \cfth). The docked molecules were the lowest-energy conformers. Thirty physisorbed or chemisorbed local minima obtained for 3P-S can be found in Fig.\,\ref{SI:3P-S_sides}, while five lowest energy structures for \cft\ and \cfth\ are depicted in Fig.\,\ref{SI:CF3-3P-S_sides} and Fig.\,\ref{SI:CF3-3P-SH_sides}, respectively. As explained in the main text, in order to consider further structural candidates for the relevant molecules in this work, namely \cft\ and \cfth, we added and substituted the necessary chemical groups to transform the docked conformers of 3P-S to the other derivatives, and reoptimized them with DFT. Only one geometry remained docked for \cfth\, and five remained docked for \cft.

The lowest energy structure of \cft\ attached to the MoS$_2$ surface (upright-standing geometry) was considered for the exploratory molecular dynamics (MD) simulation.

\clearpage
\newpage
\section{Molecular dynamics simulation}

\begin{figure}[ht]
\centering
\includegraphics[width=0.95\textwidth]{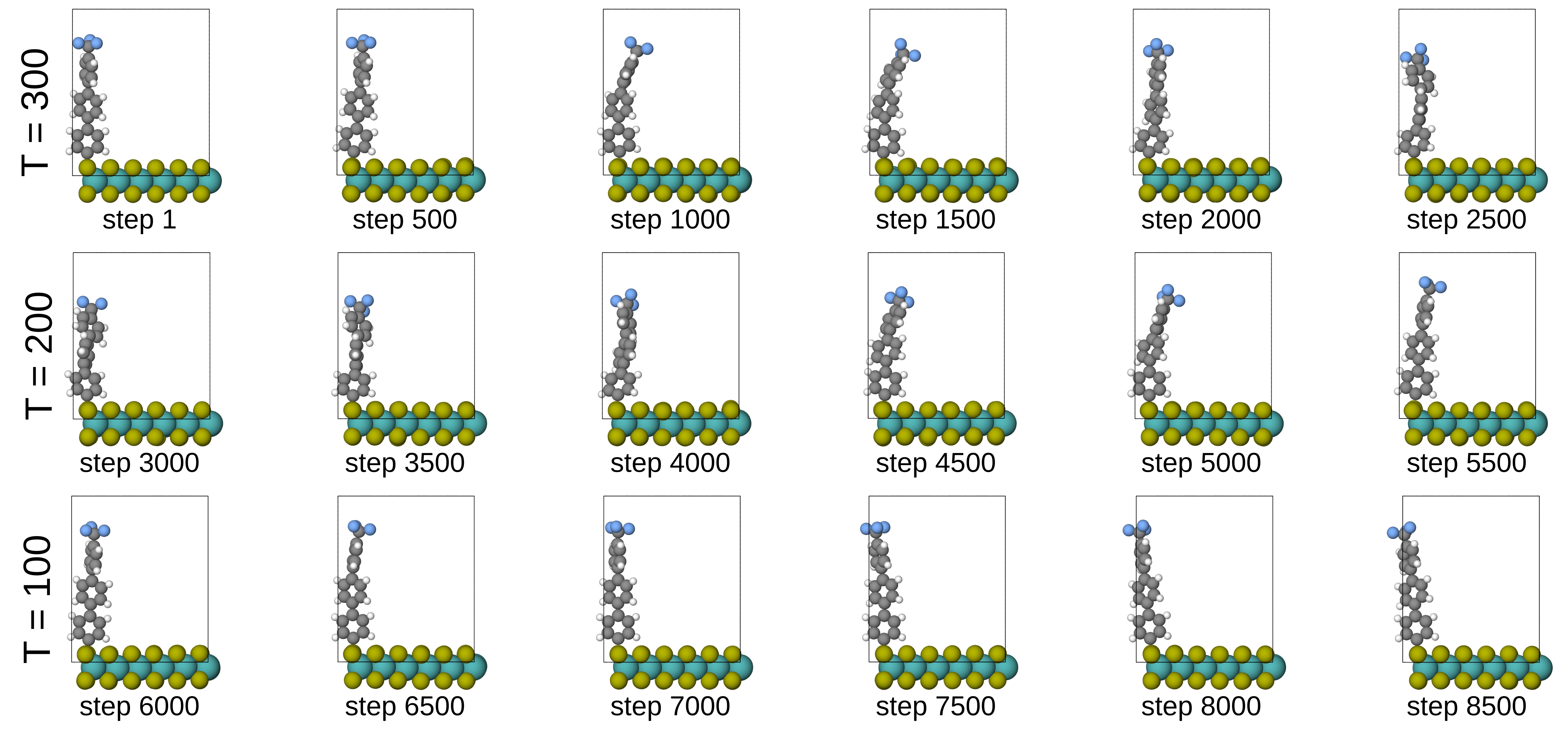}
\caption{Snapshots of molecular structures during AIMD simulation.}
\label{SI:sides}
\end{figure}

\begin{figure}[ht]
\centering
\includegraphics[width=0.99\textwidth]{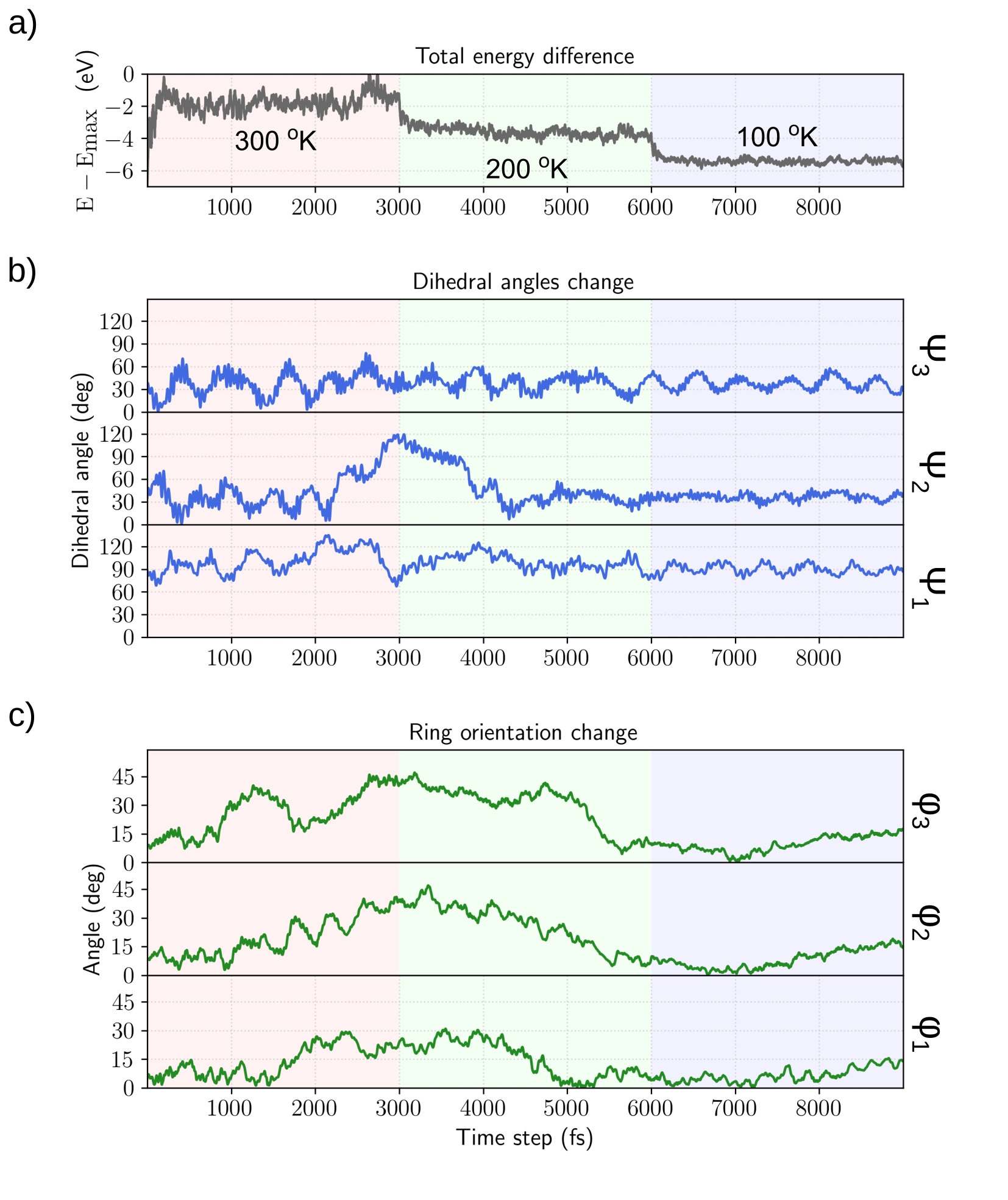}
\caption{Changes of a) total energy of the system; b) dihedral angles of the molecule; c)~ring orientation with respect to the surface of CF\textsubscript{3}-3P-S/MoS$_2$ system during the 9\,ps MD simulation.}
\label{SI:trajectory_analysis}
\end{figure}

The lowest energy configuration for \cft\ (upright-standing configuration) was used as the initial structure for exploratory MD simulations. The MD simulation was carried out for 9 ps in an NVT ensemble using a Bussi-Donadio-Parrinello thermostat \cite{10.1063/1.2408420} with 1 fs as time step and the same physical model as we used for the structure search. Starting from 300\,K after 3\,ps we decreased the temperature to 200\,K and to 100\,K after another 3\,ps. Snapshots of every 500 steps of molecular configurations during the MD simulation can be found in Fig.\,\ref{SI:sides}. Energy fluctuations during the MD simulation together with changes of dihedral angles and orientations of the molecular rings with respect to the surface denoted in Fig.\,\ref{SI:molecules_examples} can be found in Fig.\,\ref{SI:trajectory_analysis}. 

\begin{figure}[ht]
\centering
\includegraphics[width=0.95\textwidth]{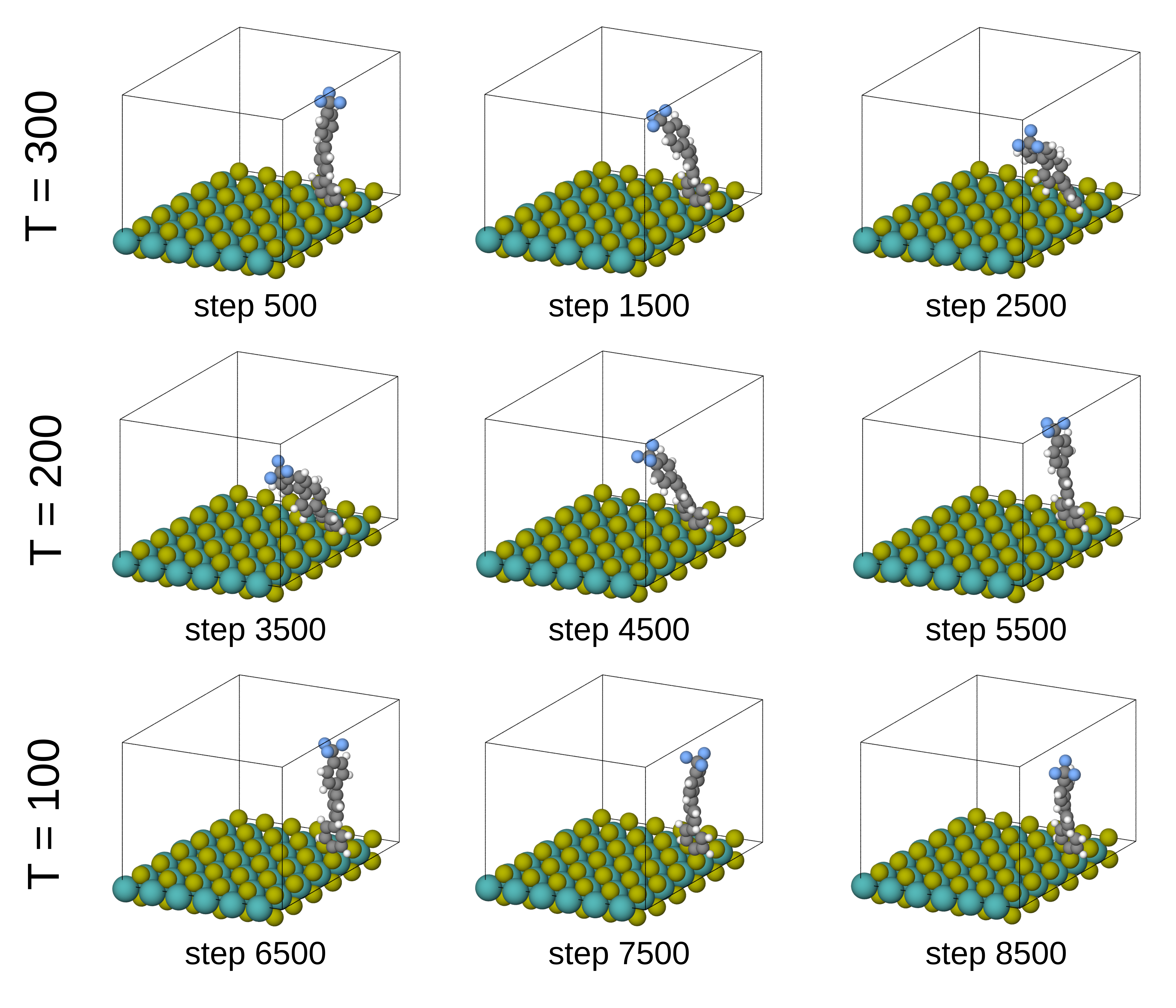}
\caption{Snapshots of molecular structures during AIMD simulation.}
\label{SI:orthos}
\end{figure}

The initial structure used for the MD simulation was the lowest energy structure of CF\textsubscript{3}-3P-S/MoS$_2$ found after geometry optimizations of structures described in the previous section. After applying a temperature of 300\,K it starts to bend and does not remain in a fully upright-standing position which can also be seen during the simulation of the molecule at 200\,K. After decreasing the temperature to 100\,K at the end of MD the molecule adopts a very similar configuration to those that was initially used as a starting configuration. 

\clearpage
\newpage

\section{Electronic Structure of Dehydrogenated \cft\ - Molecules Anchored on Neutral Vacancy}

As mentioned above and in the main text, we found five different tilt angles of \cft\ to be stable on a neutral S vacancy. For completeness, we present here the four tilt configurations that complement the case shown in the main manuscript.
For better comparison we have aligned the electronic density of states of systems with and without the vacancy or with and without the docked molecule by the energy of the $1s$ orbital of Mo. 

\begin{figure}[h]
    \centering
    \subfloat[\centering]{{\includegraphics[width=1.\linewidth]{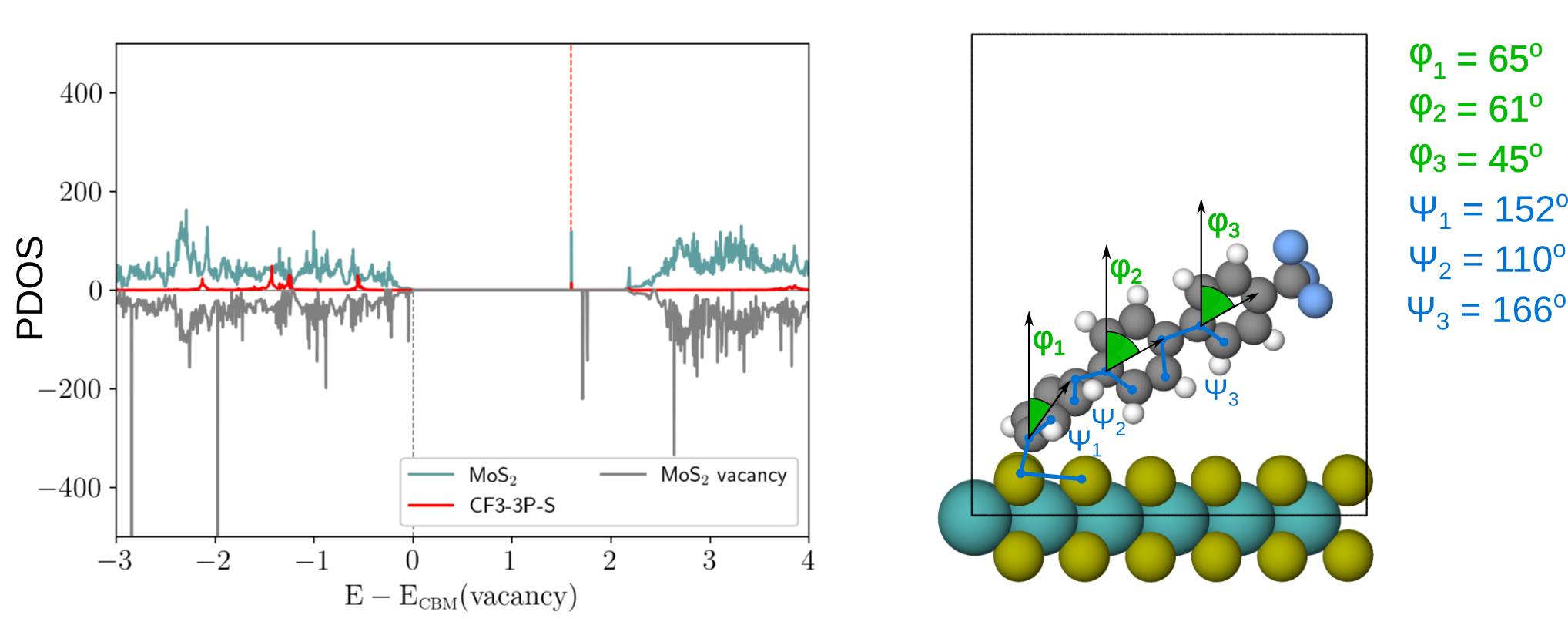} }}
       \caption{Comparison of PDOS (SOC-HSE06) for (neutral) MoS$_2$+\cft\ conformer shown on the right. The DOS of an MoS$_2$ monolayer with a vacancy is shown for comparison (grey lines). Red and grey dotted lines correspond to highest occupied states. Energy levels shifted according to the energy of the highest occupied state of MoS$_2$ with vacancy. The highest occupied state of MoS$_2$+\cft\ this geometry is singly occupied. }
  
    \label{fig:pristine_soc}
\end{figure}

\vspace{-10mm}

\begin{figure}[h]
    \centering
    \subfloat[\centering]{{\includegraphics[width=1\linewidth]{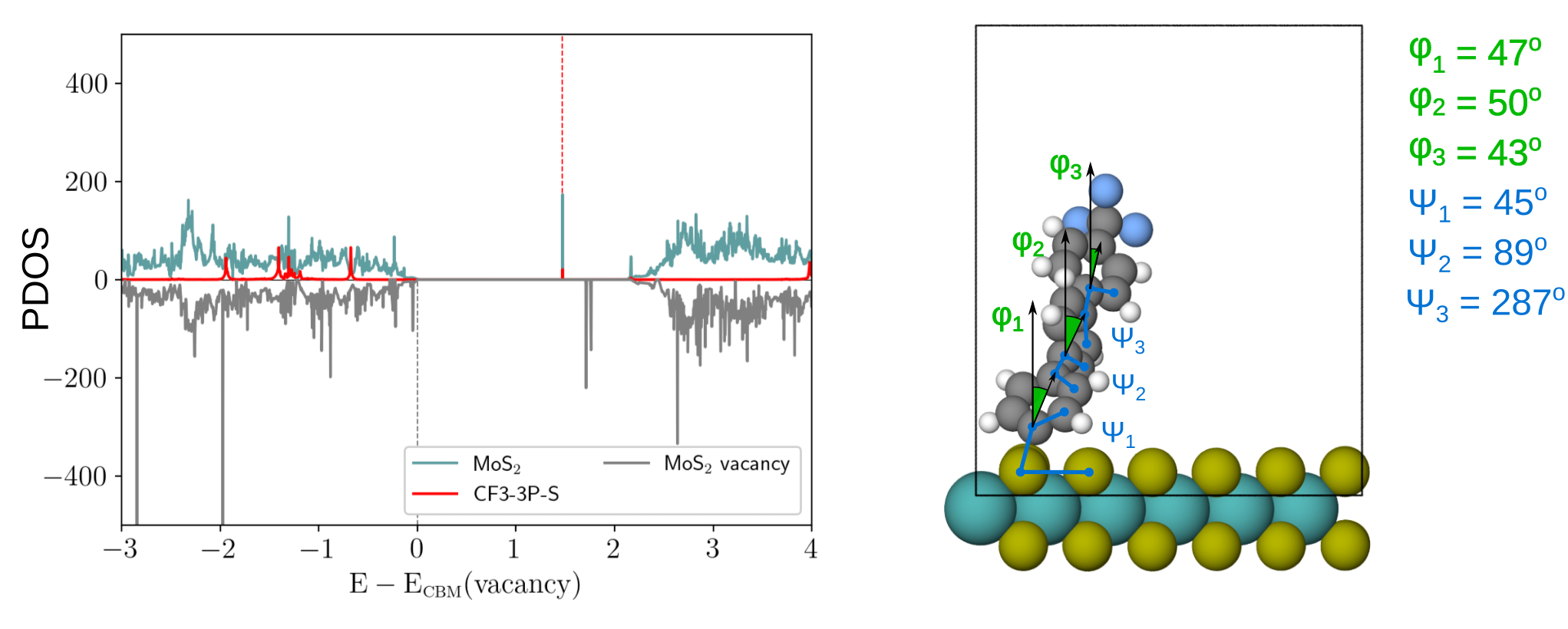} }}
      \caption{Comparison of PDOS (SOC-HSE06) for (neutral) MoS$_2$+\cft\ conformer shown on the right. The DOS of an MoS$_2$ monolayer with a vacancy is shown for comparison (grey lines). Red and grey dotted lines correspond to highest occupied states. Energy levels shifted according to the energy of the highest occupied state of MoS$_2$ with vacancy. The highest occupied state of MoS$_2$+\cft\ this geometry is singly occupied. }
     \label{fig:pristine_soc}
\end{figure}

\vspace{-10mm}

\begin{figure}[h]
    \centering
    \subfloat[\centering]{{\includegraphics[width=1\linewidth]{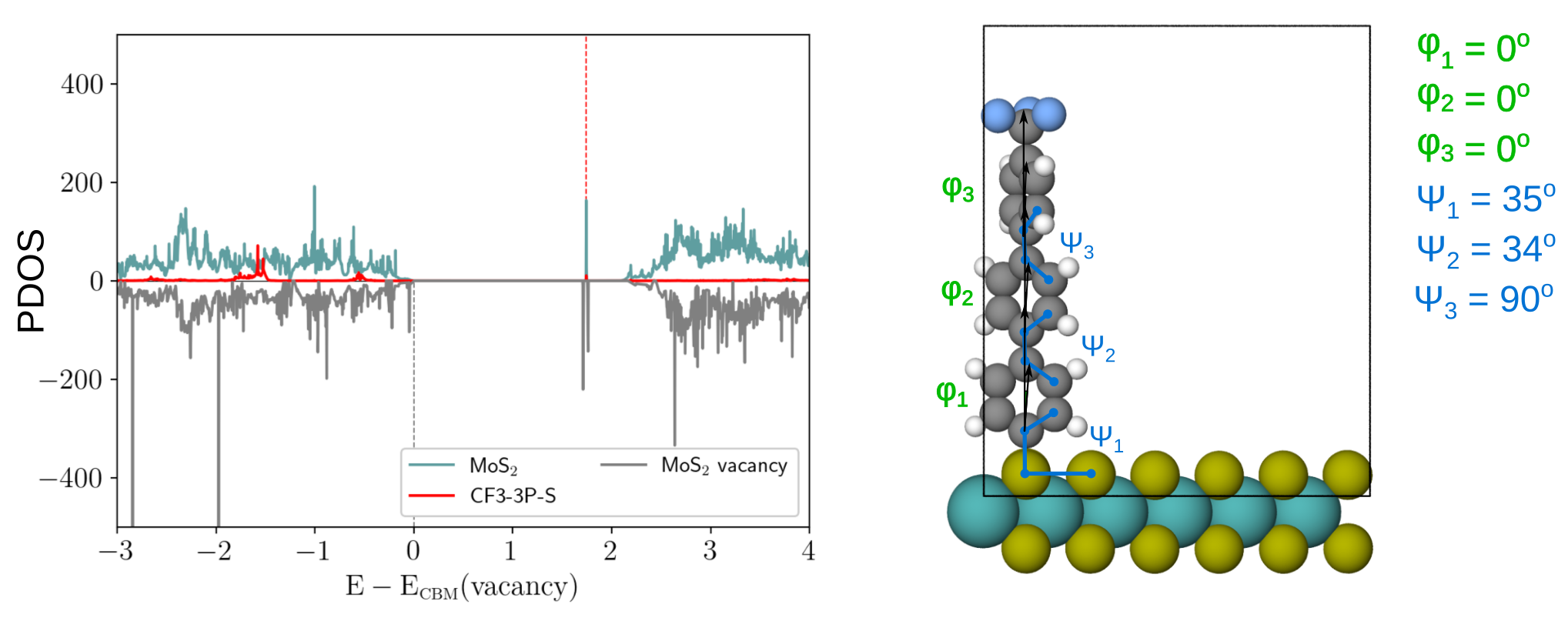} }}
     \caption{Comparison of PDOS (SOC-HSE06) for (neutral) MoS$_2$+\cft\ conformer shown on the right. The DOS of an MoS$_2$ monolayer with a vacancy is shown for comparison (grey lines). Red and grey dotted lines correspond to highest occupied states. Energy levels shifted according to the energy of the highest occupied state of MoS$_2$ with vacancy. The highest occupied state of MoS$_2$+\cft\ this geometry  is singly occupied. }
       \label{fig:pristine_soc}
\end{figure}

\vspace{-10mm}

\begin{figure}[h]
    \centering
    \subfloat[\centering]{{\includegraphics[width=1\linewidth]{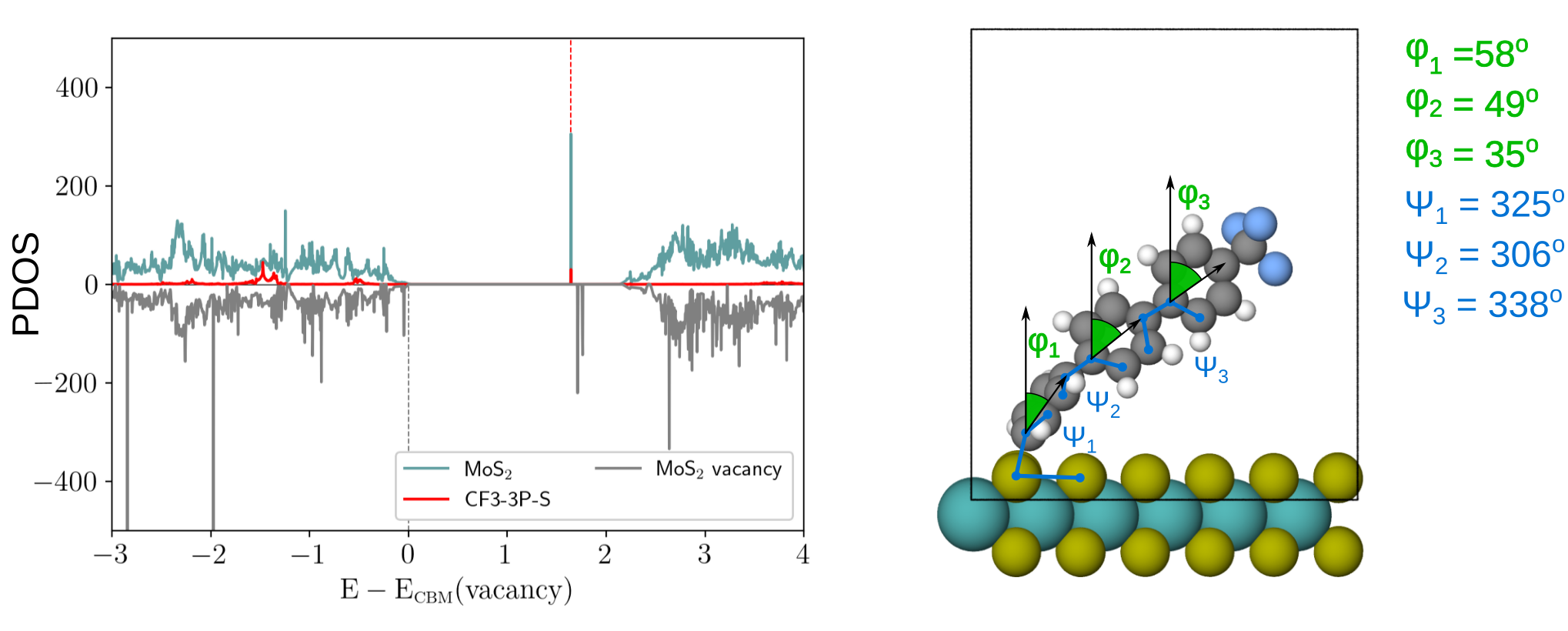} }}
      \caption{Comparison of PDOS (SOC-HSE06) for (neutral) MoS$_2$+\cft\ conformer shown on the right. The DOS of an MoS$_2$ monolayer with a vacancy is shown for comparison (grey lines). Red and grey dotted lines correspond to highest occupied states. Energy levels shifted according to the energy of the highest occupied state of MoS$_2$ with vacancy. The highest occupied state of MoS$_2$+\cft\ this geometry is \textbf{singly} occupied.} 
    \label{fig:pristine_soc}
\end{figure}

\clearpage

\clearpage
\newpage

\section{Electronic Structure of Isolated and Anchored \cft\ - Negatively Charged Vacancy}

Below we present selected simulated electronic states of isolated \cft\ and of \cft\ adsorbed on the negatively charged MoS$_2$ substrate.

 \begin{figure*}
   \centering
   \includegraphics[width=\textwidth]{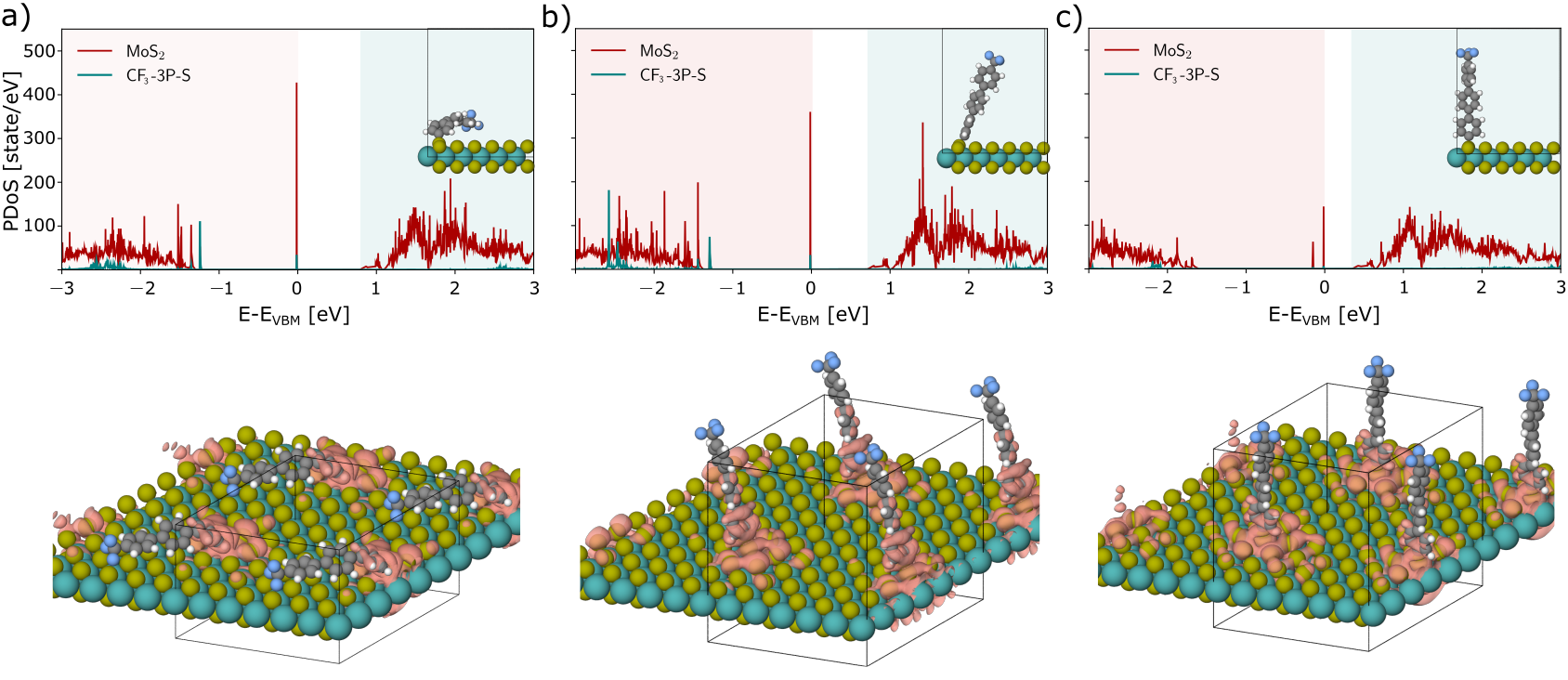}
   \caption{a) Projected electronic density of states (PDOS) of a stable conformer of \cft~anchored to the negatively-charged \mos~monolayer, obtained with density-functional theory and the HSE06 exchange-correlation functional. The molecular geometry is shown in the inset. Blue and orange lines represent the PDOS on \mos\ and on the molecule, respectively. Red shaded area marks occupied states and blue shaded area marks unoccupied states. The bottom panel shows the orbital density corresponding to the highest doubly-occupied localized state of that geometry. We show an isodensity of 0.0002 e/Bohr$^3$. Panels b) and c) show the same data as in a) for other two stable conformers.  %\\
   }
 \label{fig:theory}
 \end{figure*}

\begin{figure}[ht]
    \centering
    \includegraphics[width=0.5\textwidth]{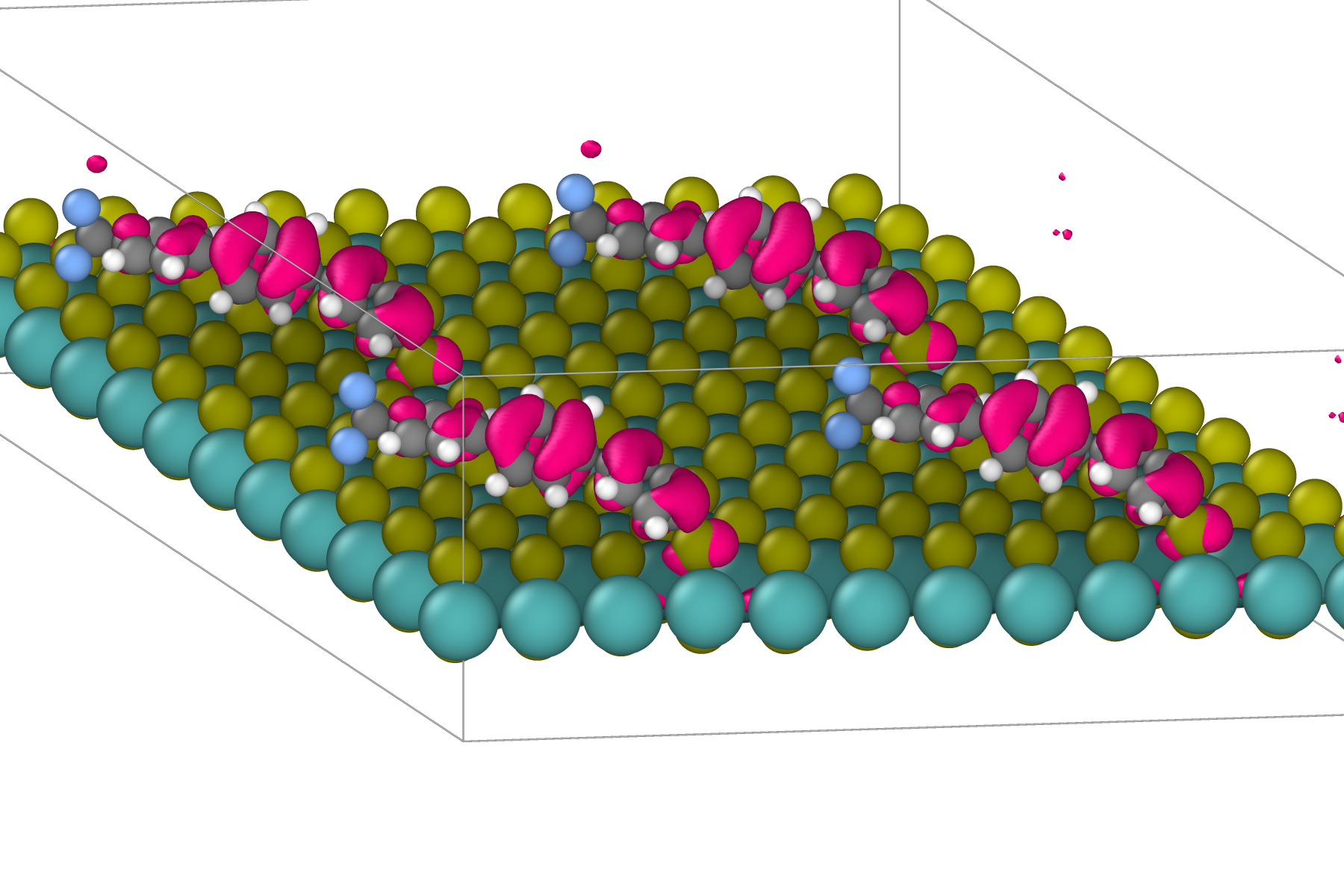}
    \caption{Orbital density (DFT-HSE06) corresponding to the
occupied molecular state at -1.2 eV in panel a of Fig.\,\ref{fig:theory}. This state has the same character as the SOMO-3 of the isolated molecule. We show an isodensity of 0.002 e/Bohr$^3$.}
    \label{fig:isolated-cft-c4}
\end{figure}

\begin{figure}[ht]
    \centering
    \includegraphics[width=0.5\textwidth]{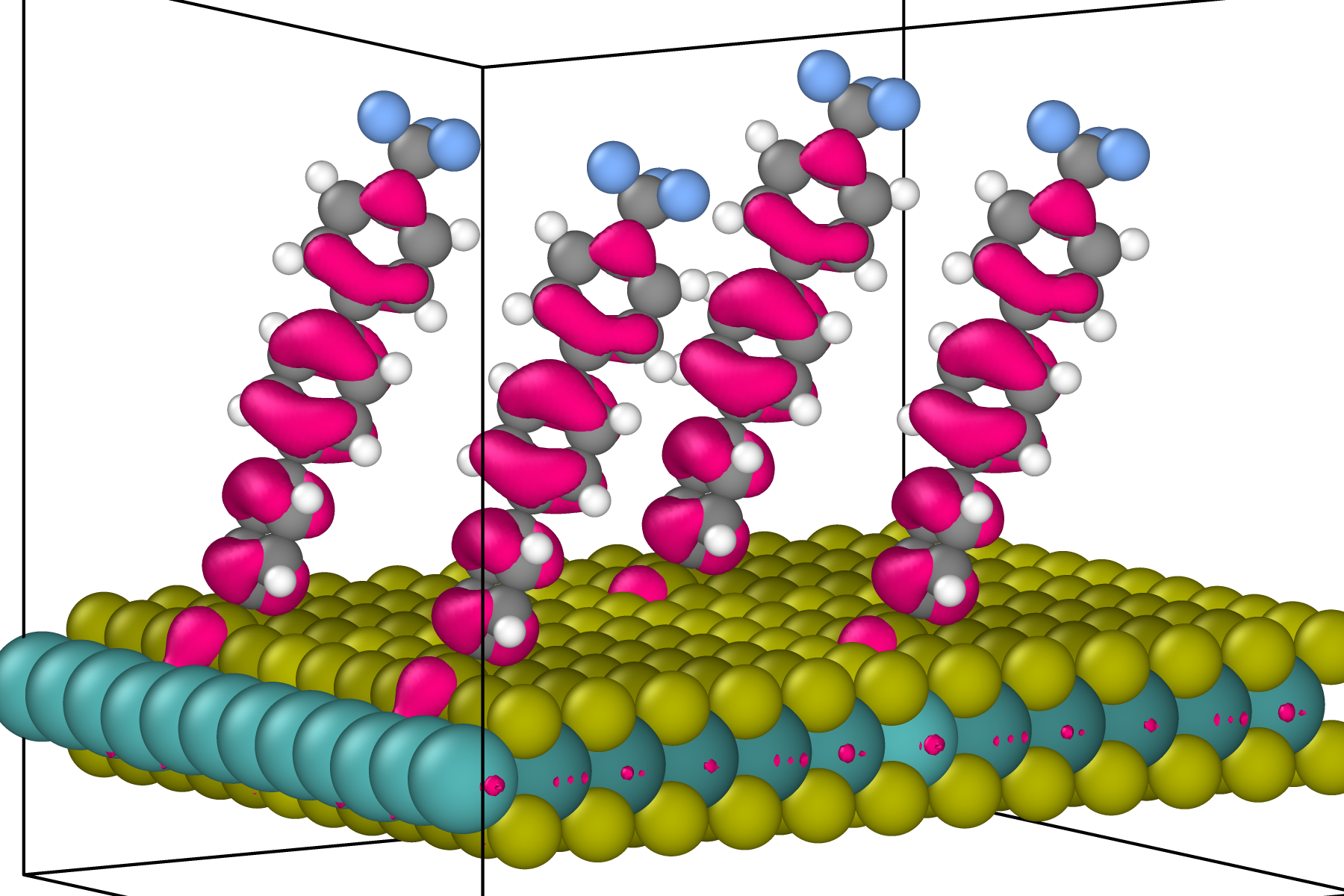}
    \caption{Orbital density (DFT-HSE06) corresponding to the
occupied molecular state at -1.3 eV in panel b of Fig.\,\ref{fig:theory} This state has the same character as the SOMO-3 of the isolated molecule. We show an isodensity of 0.002 e/Bohr$^3$.}
    \label{fig:isolated-cft-c4}
\end{figure}

\begin{figure}[ht]
    \centering
    \includegraphics[width=0.5\textwidth]{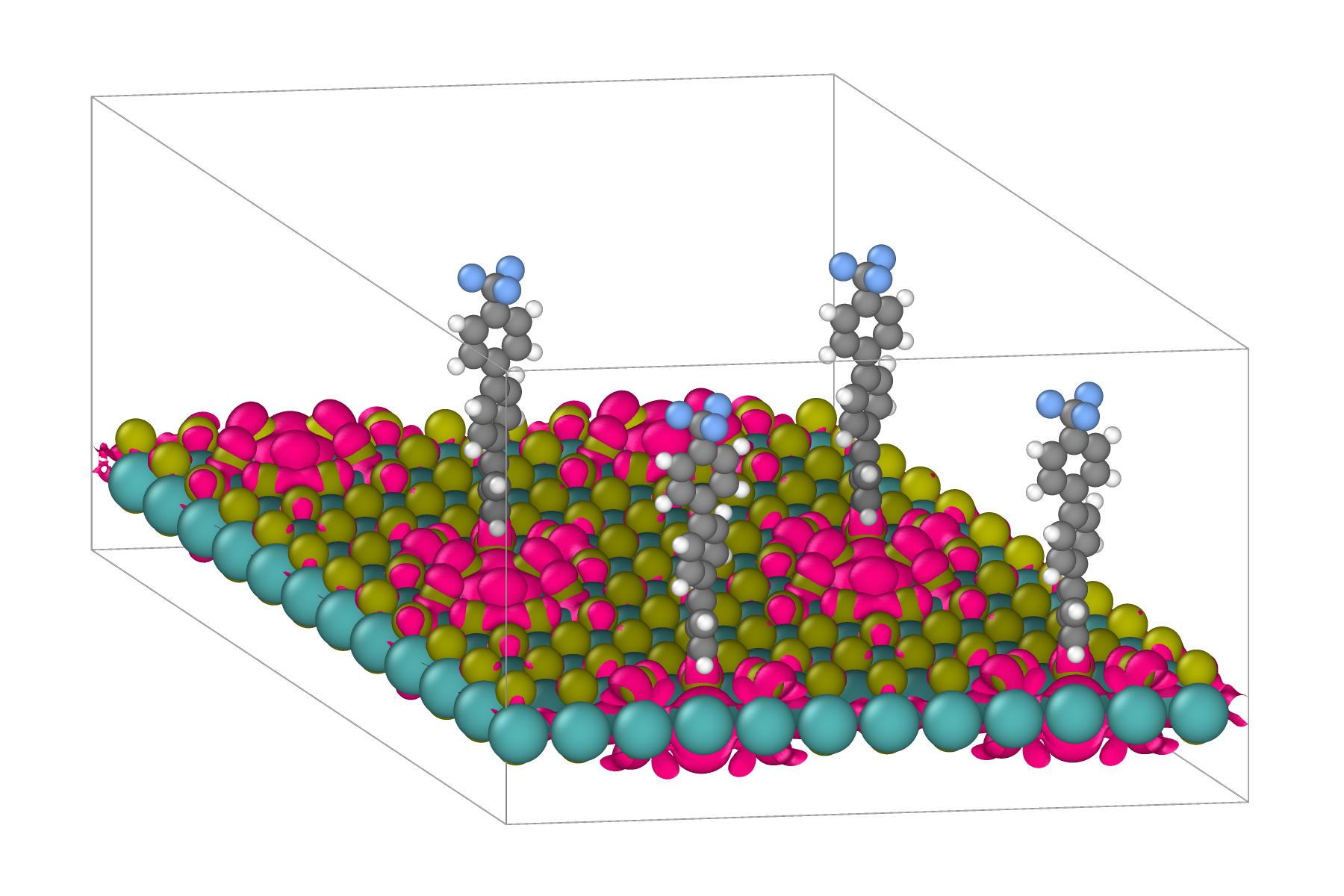}
    \caption{Orbital density (DFT-HSE06) corresponding to the
occupied molecular state at -1.6 eV in panel c of Fig.\,\ref{fig:theory}. This state has a similar character to the SOMO-1 of the isolated molecule, hybridized with the substrate. We show an isodensity of 0.0002 e/Bohr$^3$.}
    \label{fig:isolated-cft-c4}
\end{figure}

\clearpage
\newpage

\section{Variation of the electronic structure of anchored molecules depending on adsorption site}

\begin{figure}[htbp]
  \centering
\includegraphics[width=0.4\linewidth]{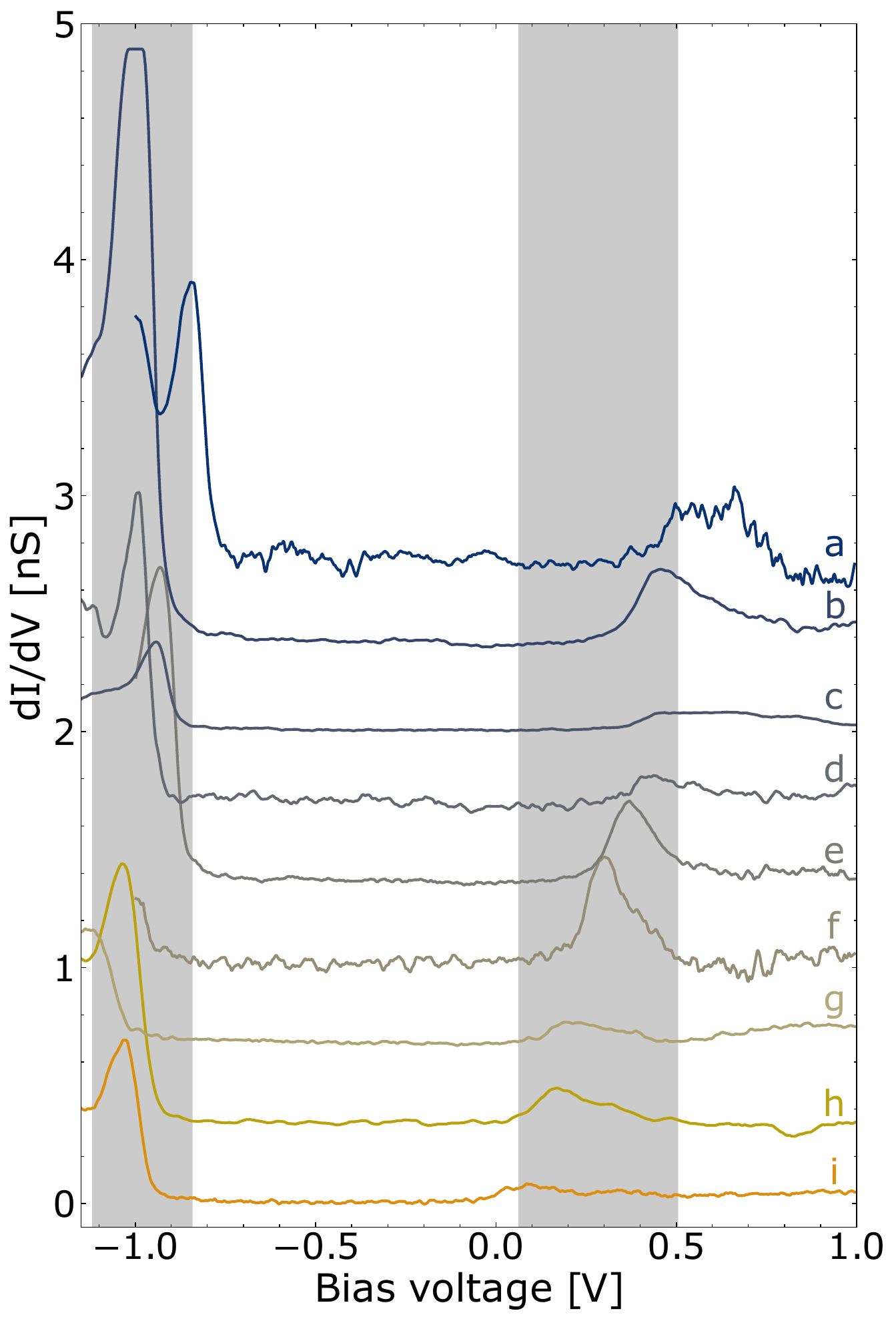}
\caption{\dIdV\ spectra of nine different anchored \cft\ molecules taken on the anchoring point, vertically offset by 0.3\,nS from each other for better visibility and sorted by energy of the onset of the NIR. The two main spectroscopic features (PIR and NIR) shift in energy, independently from each other. The grey shaded areas denote the range of this shift: For the PIR on the left the energetical position of the peak maximum is taken as the reference point, and for the NIR on the right the onset of the peaks. (Setpoints from top to bottom: 
  Feedback opened at a) 1\,V, 100\,pA,
b) 2\,V, 400\,pA,
c) 2\,V, 100\,pA, 
d) 1.5\,V, 100\,pA,
e) 1\,V, 100\,pA, 
f) 1\,V, 100\,pA, 
g) 1.5\,V, 100\,pA, 
h) 2\,V, 100\,pA, 
i) 1.5\,V, 100\,pA, 
and 2\,mV modulation amplitude for a, d and f, 5\,mV for b, e, g, h, i, and 10\,mV for c.)
}
\label{fig:spec_overview}
\end{figure}

As mentioned above, theory predicts a multitude of energetically similar states due to the rotational degree of freedom and different tilt angles of the molecules. In experiment, we observe a variation of energy-level alignments as well. The set of \dIdV spectra in Fig.\,\ref{fig:spec_overview} reflects the range of possible shifts (shaded in grey) of both the PIR and NIR resonances for different \cfth\ anchored molecules. This is likely due to a combination of molecular bending and tilt angle (as observed in theory) and additional influences, such as the moir\'e-induced modulation of the electronic structure~\cite{Trishin2021}. 

\section{Synthesis of Terphenyl-Thiols}

\subsection{Methyl(4''-(trifluoromethyl)-[1,1':4',1''-terphenyl]-4-yl)sulfane (CF\textsubscript{3}-3P-SCH\textsubscript{3})}

\begin{figure}[ht]
    \centering
    \includegraphics[width=0.5\textwidth]{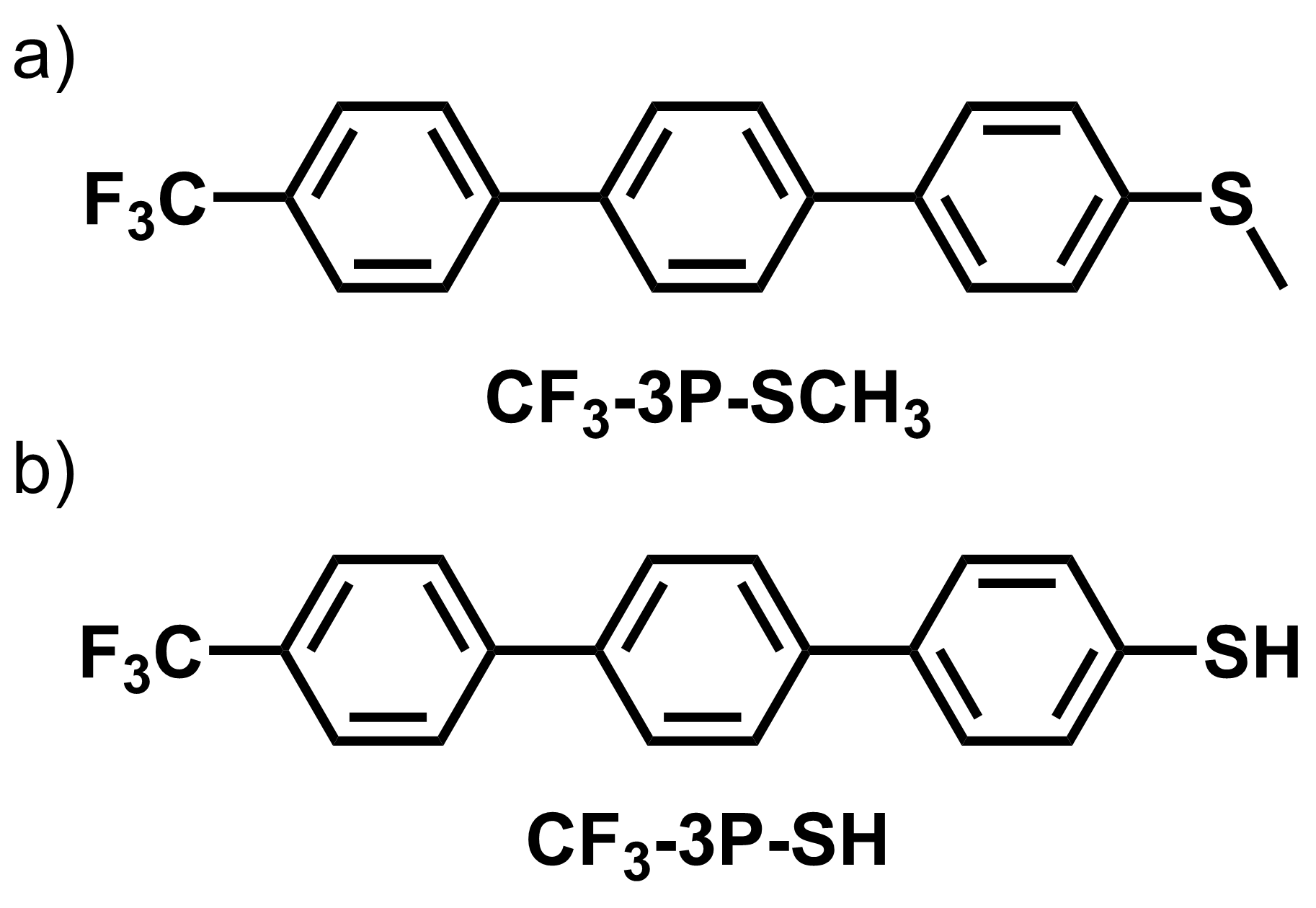}
    \caption{Structural formulae of (a)~Methyl\-(4''-(tri\-fluoro\-methyl)-\-[1,1':4',1''-\-ter\-phenyl]-4-yl)\-sulfane (\textbf{CF\textsubscript{3}-3P-SCH\textsubscript{3}}) and
    (b)~4''-(tri\-fluoro\-methyl)-[1,1':4',1''-terphenyl]-4-thiol (\textbf{CF\textsubscript{3}-3P-SH}).}
    \label{fig:synthesis}
\end{figure}

4-Bromo-4'-(tri\-fluoro\-methyl)-1,1'-biphenyl (800\,mg, 2.7\,mmol), (4-(methyl\-thio)\-phenyl)\-boronic acid (490\,mg, 2.9\,mmol), cesium carbonate (1700\,mg, 5.3\,mmol), and [1,1'-bis(di\-phenyl\-phos\-phino)\-ferro\-cene]\-di\-chloro\-palla\-dium(II), complex with di\-chloro\-me\-thane (0.08\,mmol, 70\,mg) were kept under vacuum in a Schlenk tube for 15\,min. A mixture of dry tetra\-hydofuran (12\,mL), dry ethanol (6\,mL), and dry \textit{N,N}-di\-methyl\-formamide (3\,mL) was added to the flask. After subsequently applying vacuum and argon for several times, the mixture was stirred under argon over night at 80°C. The mixture was allowed to cool to room temperature and water and dichloro\-methane were added. Phases were separated and the aqueous phase was extracted with dichloro\-methane two more times. The combined organic phases were dried with magnesium sulfate and the solvent was evaporated. The crude product was purified by column chromatography (petroleum ether/dichloro\-methane 9:1 to 7:3) to afford 790 mg (2.3\,mmol, 86\% yield) of terphenyl \textbf{CF\textsubscript{3}-3P-SCH\textsubscript{3}} (see Fig.\,\ref{fig:synthesis}a).

\textsuperscript{1}H-NMR (500\,MHz, CDCl\textsubscript{3}): $\delta$[ppm] = 7.76 - 7.70 (m, 4H), 7.70 - 7.65 (m, 4H), 7.60 - 7.55 (m, 2H), 7.38 - 7.33 (m, 2H), 2.54 (s, 3H).

\textsuperscript{13}C-NMR\footnote{Due to low intensity and C-F coupling, \textsuperscript{13}C peaks for \underline{C}F\textsubscript{3} and the adjacent quarternary carbon were nor found or not assigned.} (126\,MHz, CDCl\textsubscript{3}): $\delta$[ppm] = 144.3, 140.6, 138.7, 138.3, 137.3, 127.8, 127.5, 127.5, 127.4, 127.1, 125.9 (q, $J$ = 3.2\,Hz), 16.0.

HRMS (ESI\textsuperscript{+}): $m/z$ = [M]\textsuperscript{+} 344.0843 (calc. C\textsubscript{20}H\textsubscript{15}F\textsubscript{3}S\textsuperscript{+} = 344.0842).

\subsection{4''-(trifluoromethyl)-[1,1':4',1''-terphenyl]-4-thiol (CF\textsubscript{3}-3P-SH)}

A mixture of \textbf{CF\textsubscript{3}-3P-SCH\textsubscript{3}} (150\,mg, 0.43\,mmol), and sodium ethanethiolate (360\,mg, 4.3\,mmol) in 8\,mL of $N,N$-dimethylformamide was degassed by applying an argon flow for 5\,min. The reaction was stirred for 2h under argon at 150°C. After cooling to room temperature, 20\,mL of water and 7\,mL of hydrochloric acid (1\,mol/L) were added. The colorless precipitate was separated by vacuum-filtration and subsequently washed with water. The product was collected and dried under vacuum at 60°C to give 125\,mg (0.38\,mmol, 88\% yield) of \textbf{CF\textsubscript{3}-3P-SH}  (see Fig.\,\ref{fig:synthesis}b).

\textsuperscript{1}H-NMR (500\,MHz, CDCl\textsubscript{3}): $\delta$[ppm] = 7.76 - 7.69 (m, 4H), 7.70 - 7.65 (m, 4H), 7.55 - 7.49 (m, 2H), 7.41 - 7.34 (m, 2H), 3.52 (s, 1H).

\textsuperscript{13}C-NMR\footnote{Due to low intensity and C-F coupling, \textsuperscript{13}C peaks for \underline{C}F\textsubscript{3} and the adjacent quarternary carbon were nor found or not assigned.} (126\,MHz, CDCl\textsubscript{3}): $\delta$[ppm] = 144.3, 140.3, 138.8, 138.0, 130.5, 130.0, 127.9, 127.8, 127.5, 127.4, 125.93 (q, $J$ = 3.6\,Hz).

HRMS (ESI\textsuperscript{-}): m/z = [M-H]\textsuperscript{-} 329.0614 (calc. C\textsubscript{19}H\textsubscript{12}F\textsubscript{3}S\textsuperscript{-} = 329.0617)

For the application in UHV/for STM, \textbf{CF\textsubscript{3}-3P-SH} was further purified by temperature gradient sublimation.

\clearpage

\bibliography{bibliography}